\documentclass[%
 aip,
 amsmath,amssymb,
 reprint,%
]{revtex4-1}

\usepackage{graphicx}
\usepackage{dcolumn}
\usepackage{bm}
\usepackage{nicefrac}
\usepackage[utf8]{inputenc}
\usepackage[T1]{fontenc}
\usepackage{mathptmx}

\begin{document}

\preprint{AIP/123-QED}

\title[]{Pressure, temperature, and orientation dependent thermal conductivity \\of $\alpha$-1,3,5-trinitro-1,3,5-triazinane ($\alpha$-RDX)}

\author{Romain Perriot}
\email{rperriot@lanl.gov}
\affiliation{Theoretical Division, Los Alamos National Laboratory, Los Alamos, NM 87545, USA}

\author{Michael S. Powell}
\affiliation{High Explosives Science and Technology, Los Alamos National Laboratory, Los Alamos, NM, 87545, USA}

\author{John D. Lazarz}
\affiliation{Shock and Detonation Physics, Los Alamos National Laboratory, Los Alamos, NM, 87545, USA}

\author{C. A. Bolme}
\affiliation{Shock and Detonation Physics, Los Alamos National Laboratory, Los Alamos, NM, 87545, USA}

\author{Shawn D. McGrane}
\affiliation{Shock and Detonation Physics, Los Alamos National Laboratory, Los Alamos, NM, 87545, USA}

\author{David S. Moore}
\affiliation{Shock and Detonation Physics, Los Alamos National Laboratory, Los Alamos, NM, 87545, USA}

\author{M. J. Cawkwell}
\affiliation{Theoretical Division, Los Alamos National Laboratory, Los Alamos, NM 87545, USA}

\author{Kyle J. Ramos}
\affiliation{High Explosives Science and Technology, Los Alamos National Laboratory, Los Alamos, NM, 87545, USA}

\date{\today}

\begin{abstract}
We use reverse non-equilibrium molecular dynamics ($R$NEMD) simulations to determine the thermal conductivity in $\alpha$-RDX in the $<$100$>$, $<$010$>$, and $<$001$>$ crystallographic directions. Simulations are carried out with the Smith-Bharadwaj non-reactive empirical interatomic potential [Smith \& Bharadwaj, J. Phys. Chem. B 103, 3570 (1999)], which represents the thermo-elastic properties of RDX with good accuracy.  As an illustration, we report the temperature and pressure dependence of lattice constants of $\alpha$-RDX, which compare well with experimental and $ab\ initio$ results, as do linear and volume thermal expansion coefficients, which we also calculate. We find that the thermal conductivity depends linearly on the inverse temperature in the 200-400~K regime due to the decrease in the phonon mean free path. The thermal conductivity also exhibits anisotropy, with a maximum difference at 300~K of 24\% between the $<$001$>$ and $<$010$>$ directions, an effect that remains when temperature increases. Thermal conductivity in the $<$100$>$ direction is mostly between the two other directions, although crossovers are predicted with $<$001$>$ at high temperature, and $<$010$>$ at low temperature under pressure. We observe that the thermal conductivity varies linearly with pressure up to 4~GPa. The data are fitted to analytical functions for interpolation/extrapolation and use in continuum simulations. MD results are validated against experiments using impulsive stimulated thermal scattering (ISTS) on RDX single crystals at 293~K and ambient pressure, showing good qualitative and quantitative agreement: same ordering between the three principal orientations, and an average error of 10\% between the experiments and the model. These results provide confidence that the extracted analytical functions using the $R$NEMD methodology and the Smith-Bharadwaj potential can be applied to model the thermal conductivity of $\alpha$-RDX. 
\end{abstract}

\maketitle

\section{Introduction}

The accidental initiation of an explosive material subject to an insult is a major concern to both manufacturers and users.~\cite{Bowden} In order to mitigate this type of risk, substantial effort is devoted to predicting the response of high explosives (HE) to weak shock stimuli, efforts which require accurate models that notably account for the role of the material's microstructure. Indeed, most explosives are characterized by a non-uniform microstructure, where internal interfaces play critical roles in the reaction process.~\cite{Storm1989} 
While the length scale associated with these HE characteristics (ten nm to a few $\mu$m) invites the use of mesoscale models, those require accurate parametrizations that are usually obtained from smaller scale simulations (i.e. atomistic).

One of the key inputs to a mesoscale model is the thermal conductivity of the material, because heat localization is a necessary precursor to chemical reactions.~\cite{Bowden} Anisotropy, pressure, and temperature, are effects that all must be introduced in order to yield meaningful predictions of the material's behavior. Experimentally,~\cite{McGuire1981,Fedorov1960,Rogers1975,Loftus1959,Dobratz1985,Zinn1960,Miller1997,Faubion1976,Shoemaker1985,Hanson-Parr1999, Lawless2020} most studies of the thermal conductivity of explosives used polycrystalline and sometimes porous samples, such that detailed information on orientation dependences are not readily available. However, since bulk thermal transport is an atomistic characteristic,~\cite{Kittel} molecular dynamics (MD) simulations are well suited to provide insights into the thermal properties that inform mesoscale models. Classical MD with empirical force fields was for instance used on two HE crystals, to highlight the anisotropy and the impact of defects on the thermal conductivity of 2,4,6-triamino-1,3,5-trinitrobenzene (TATB),~\cite{Kroonblawd2016,Kroonblawd2016b} and the thermal conductivity in liquid and crystalline 1,3,5,7-tetranitro-1,3,5,7-tetrazoctane (HMX).~\cite{Bedrov2000,Chitsazi2020}

In this work, we present results from MD calculations that predict the thermal conductivity, $\kappa$, of the heavily used secondary explosive 1,3,5-trinitro-1,3,5-triazinane (RDX). The effects of the crystal orientation, temperature, and pressure were determined by using the $reverse$ non-equilibrium molecular dynamics ($R$NEMD) technique,~\cite{Muller-Plathe1997, Zhang2005} with an accurate non-reactive empirical interatomic potential that has been shown to reproduce well the thermomechanical response of RDX.~\cite{Munday2011, Josyula2014,Hooks2015} We focused on the $\alpha$ polymorph of RDX, which is stable under ambient conditions and up to $\sim$4~GPa, above which the $\gamma$ phase is favored.\cite{Olinger1978, Davidson2008, Dreger2012, Cawkwell2016} We find mostly linear responses to pressure and the inverse temperature in the regime considered here, corresponding to moderate insults and operating conditions (200--400~K, 0--4~GPa). We also find that $\alpha$--RDX displays a notable anisotropy between the $<$100$>$, $<$001$>$, and $<$001$>$ directions (24\% between $<$001$>$ and $<$010$>$ at 300~K and 0~GPa). Analytical expressions for $\kappa$(T,P) are extracted for each orientation, allowing for the full determination of the crystal thermal conductivity. Additionally, we show results from impulsive stimulated thermal scattering (ISTS) to validate the $R$NEMD results. This is a critical contribution, since, as mentioned above, most experimental results use polycrystalline samples with varying porosity, which prevents the direct comparison of MD simulations and experiments. The results, which were obtained using oriented single crystals, agree quantitatively with the $R$NEMD. 

The paper is organized as follows: computational details about the empirical potential used in the study, the $R$NEMD technique, and simulation geometry preparation, and the ISTS methodology and setup are given in section~\ref{sec:methods}; the lattice constants and linear and volume thermal expansion coefficients obtained from the MD simulations are discussed in section~\ref{sec:lattice}; the effect of the temperature on the thermal conductivity is discussed in section~\ref{sec:T}, while the effect of pressure on the thermal conductivity is presented in the next section (\ref{sec:P}). In section~\ref{sec:final}, we provide a model and coefficients for $\kappa(T,P)$ for the three principal directions of the crystal. ISTS results are presented in section~\ref{sec:ISTS}. Discussion (sec.~\ref{sec:discussion}) and conclusions (sec.~\ref{sec:conclusions}) follow.

\section{Methods}\label{sec:methods}
\subsection{Smith-Bharadwaj empirical potential}
The MD simulations are performed with the Smith-Bharadwaj (SB) empirical potential (SB-FF) for nitramine explosives that was parameterized to high level quantum chemical calculations.~\cite{Smith1999}   
It is a non-reactive force field, where the total energy is the sum of five terms:
\begin{itemize}
\item bond stretches: $U_{1}=\frac{1}{2}k^s_{ab}(r_{ij}-r_{ij}^0)^2$;
\item valence bends: $U_{2}=\frac{1}{2}k^b_{abc}(\theta_{ijk}-\theta_{ij}^0)^2$;
\item torsions: $U_{3}=\frac{1}{2}k^t_{abcd}\left[1-\textrm{cos}(n\phi_{ijk})\right]$;
\item out-of-plane bends: $U_{4}=\frac{1}{2}k^d_{abcd}\delta_{ijkl}^2$;
\item nonbonded: $U_{5}=A_{ab}\times\textrm{exp}(-B_{ab}r_{ij})-\frac{C_{ab}}{r_{ij}^6} + \frac{q_iq_j}{4\pi\epsilon_0r_{rij}}$;
\end{itemize}

\noindent where $k^s$, $k^b$, $k^t$, and $k^d$ are force constants, $r$, interatomic distances, $\theta$, $\phi$, and $\delta$ are angles (dihedral and out-of-plane for the latter two), and $A$, $B$ and $C$ are constants for the Buckingham-type non-bonded interactions; indices $a,b,c,d$ refer to the atom type of atoms $i,j,k,l$, respectively. For the value of each parameter, the reader is referred to the original paper by Smith and Bharadwaj.~\cite{Smith1999} Electrostatic interactions are calculated with the modified set of partial charges $q_i$ proposed by Bedrov et al..~\cite{Bedrov2001}
 
The force field was successfully applied to reproduce the unit cell parameters, coefficients of thermal expansion, and heat of sublimation of several HMX polymorphs,~\cite{Bedrov2001} and was also used to predict the thermal conductivity of HMX.~\cite{Bedrov2000, Chitsazi2020} The SB-FF was also applied to RDX, and shown to reproduce lattice and elastic constants, and the equation of state of the $\alpha$ and $\gamma$ polymorphs.~\cite{Munday2011,Josyula2014,Hooks2015} Beyond this, studies focused on the behavior of $\alpha$-RDX under indentation,~\cite{Weingarten2015} the mechanisms of the $\alpha\rightarrow\gamma$ phase transformation,~\cite{Josyula2019} and the response of RDX to shock compression,~\cite{Cawkwell2010, Ramos2010, Bidault2018} were also performed with the SB-FF. The good performance of the SB-FF for RDX is not too surprising, considering its accuracy for HMX, for which it was designed, and the similarities between the two nitramine molecules.


\subsection{$R$NEMD simulations}

The thermal conductivity $\kappa$ of a material is defined by Fourier's law as:

\begin{equation}\label{eq:fourier}
\kappa=\frac{\boldsymbol{J}}{\nabla T},
\end{equation} 

\noindent where $\boldsymbol{J}$ is the heat flux response to a temperature gradient $\nabla T$, and vice versa. Following, there are two main paths to calculate $\kappa$ from MD simulations: 

1) $Impose$ $\nabla T$, $measure$ $\boldsymbol{J}$ \newline
This is similar to an experimental protocol, and is commonly referred to as the ``direct'' method; the temperature gradient is imposed by thermostating distant regions of the sample at different temperatures (namely, $T_{hot}$ and $T_{cold}$), and measuring the response heat flux $\boldsymbol{J}$; the latter can be done by monitoring the thermostats and extracting the amount of energy added/subtracted from the respective regions. While the direct method is intuitive and straightforward to implement in simulations, it suffers from serious practical difficulties, as was discussed in Refs.~\onlinecite{Muller-Plathe1997} and~\onlinecite{Schelling2002}: $\boldsymbol{J}$ is a quantity with large fluctuations, such that large temperature gradients must be used in order to distinguish trend from noise. These temperature gradients are also usually out of the range of experimental measurements, such that the results might not be directly comparable to experiments. Additionally, a large $\nabla T$ leads to non-linearities near the thermostated regions, which violates the linear response assumption governing Eq.~\ref{eq:fourier}.

2) $Impose$ $\boldsymbol{J}$, $measure$ $\nabla T$ \newline
An alternative to the direct method was proposed by M\"uller-Plathe:~\cite{Muller-Plathe1997, Zhang2005} since $\boldsymbol{J}$ is the problematic quantity, one can, by contrast with the direct method, impose the flux, which is therefore known exactly.  Further, the temperature gradient $\nabla T$ becomes the response quantity. Since $\nabla T$ is measured over a reasonable number of particles, it results in much better statistical averages. The heat flux is obtained by periodically swapping molecular center-of-mass velocities of ``hot'' and ``cold'' molecules between ``cold'' and ``hot'' regions, in order to make the ``cold'' region colder, and the ``hot'' region hotter. This is known as the $reverse$ non-equilibrium molecular dynamics ($R$NEMD), or M\"uller-Plathe, technique, and is the method employed in this paper. The $R$NEMD method has been validated against both the direct method and the more computationally extensive and explicit Green-Kubo method.~\cite{Muller-Plathe1997, Schelling2002, Izvekov2011}

\begin{figure}
\centering
\includegraphics[width=1\columnwidth]{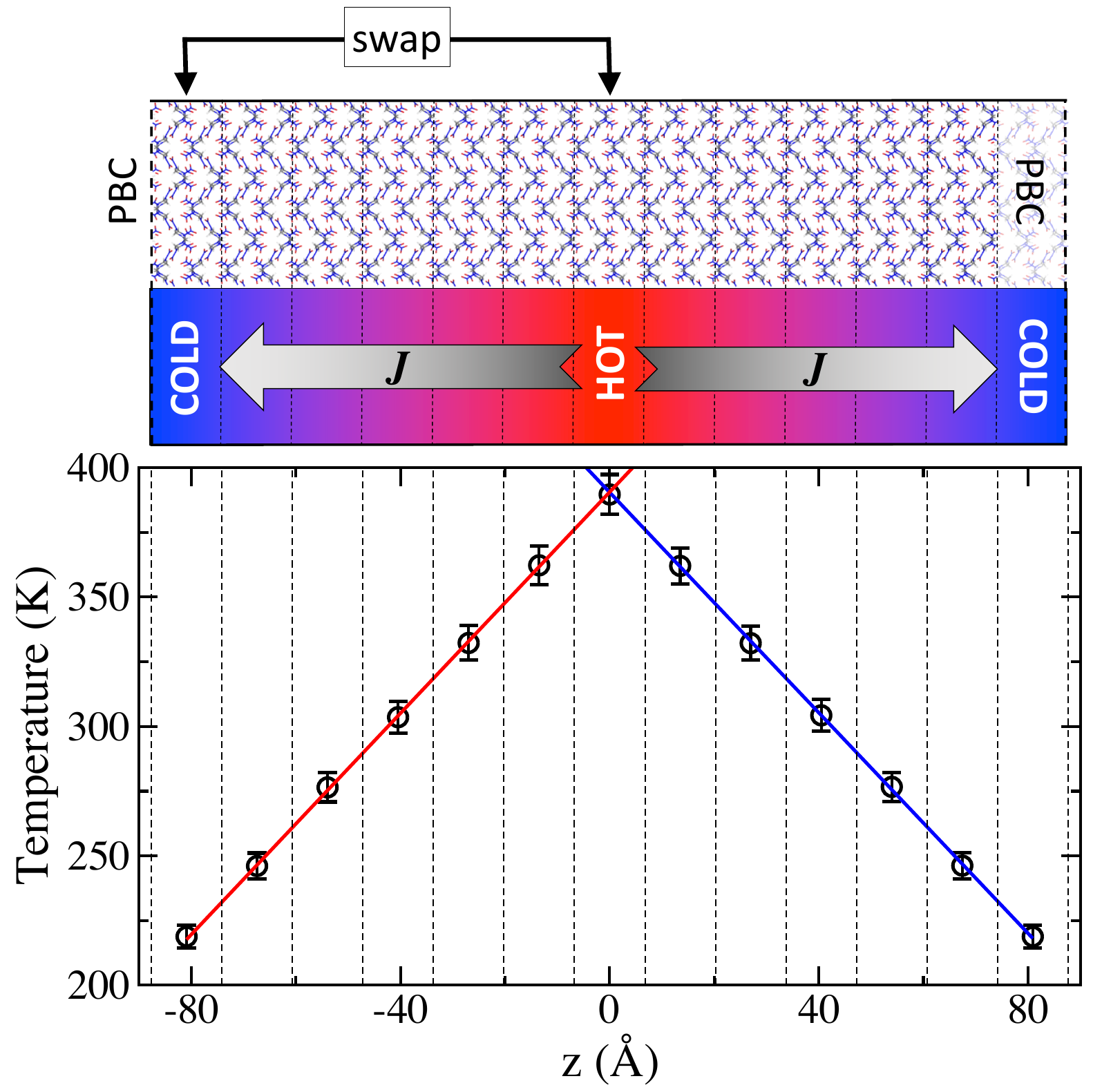}
\caption{\label{fig:RNEMD}$R$NEMD scheme. Here, the sample consists of 12$\times$3$\times$3 unit cells oriented in the $<$100$>$ direction. Each layer consist of 1$\times$3$\times$3 unit cells and PBC are applied on all directions, such that the cold layer is repeated in the figure. Velocity swaps occur between the ``cold'' and ``hot'' regions, producing a heat flux $\boldsymbol{J}$ away from the hot region and a resulting temperature gradient $\nabla T$ that is extracted from the temperature profile. In this particular case ($T$=300~K, $P$=0~GPa), we obtained $T(z)=390.4-2.13*|z|$ and $T(z)=390.6-2.14*|z|$ with the origin in the middle of the hot ``region''. }
\end{figure}

In detail, the $R$NEMD implementation is as follow (see also Fig.~\ref{fig:RNEMD}):
\begin{enumerate}
\item Equilibrate the system at the target average temperature, using for instance a Langevin thermostat (NVT simulations).
\item Define hot and cold regions and intermediate layers, and identify which molecules belong to each region. 
\item Remove the thermostat and evolve the whole system in the constant energy ensemble (NVE) for a time $t$.
\item Identify the molecule in the hot region with the smallest kinetic energy, and the molecule in the cold region with the largest kinetic energy.
\item Exchange the atomic velocities of the two molecules.
\item Repeat N times from step 3. 
\end{enumerate}

\begin{figure}
\centering
\includegraphics[width=0.9\columnwidth]{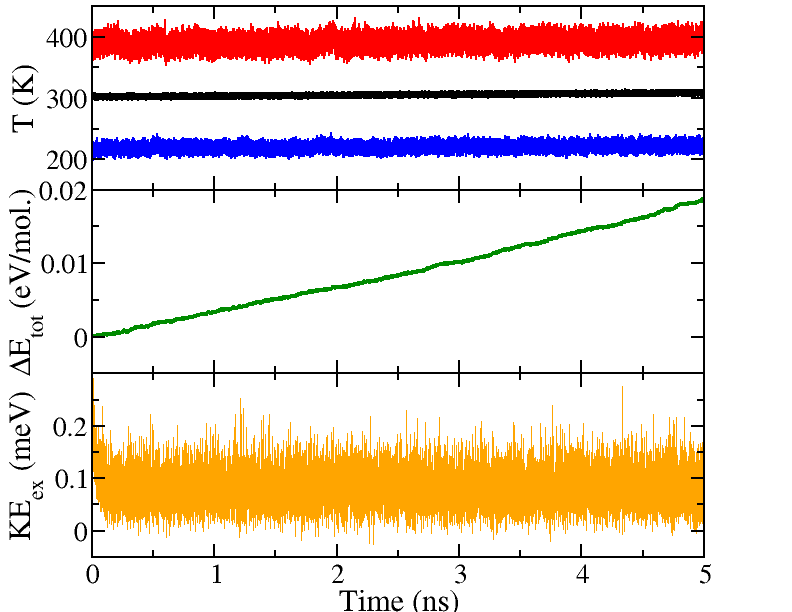}
\caption{\label{fig:converge}Profiles from the $R$NEMD simulation for $\alpha$-RDX at 300~K, 0~GPa, in the $<$100$>$ orientation (12$\times$3$\times$3 unit cells). Top: temperature profiles for the hot region (red), cold region (blue), and sample (black). Middle: total energy profile. Bottom: kinetic energy exchanged at each swap. The total simulation time of 5~ns corresponds to 20000~swaps.}
\end{figure}

\noindent The MD simulations (steps 1 and 3) were performed with the LAMMPS code;~\cite{Plimpton1995} while the identification and swapping procedures, as well as the processing of inputs/outputs were performed with in-house tools. At each swap of velocities (step 5), the exact amount of heat exchanged is known, and the flux along the direction $\alpha$ is given by:

\begin{equation}\label{eq:J}
\boldsymbol{J}_\alpha=\frac{\sum_{\textrm{transfers}}\frac{m}{2}(v_h^2-v_c^2)}{2\times A\times t},
\end{equation}

\noindent where $m$ is the mass of the molecule, $v_h$ and $v_c$ the center-of-mass velocities of the hot and cold molecules, respectively, $A$ the cross-sectional area of the sample, and t the total time of the simulation. The factor of 2 accounts for the two directions of the heat transfer (left and right, see Fig.~\ref{fig:RNEMD}). Thermal equilibrium is reached after a number of cycles, see Fig.~\ref{fig:converge}. The temperature gradient is then calculated by performing a linear fit of the temperature as a function of the distance from the hot region (Fig.~\ref{fig:RNEMD}). The average temperature in each layer is obtained according to $T=\nicefrac{2E_{\textrm{kin}}}{3Nk_B}$, with $E_{\textrm{kin}}$ the total kinetic energy in the layer, $N$ the number of atoms in the layer, and $k_B$ the Boltzmann constant. 

From Fig.~\ref{fig:converge}, we notice that the total energy exhibits a nearly constant but very small drift during the course of the simulation. The origin of the drift is unknown. Simple NVE simulations with an identical system showed similar behavior, such that the swapping procedure is not at fault; the effect also remains when a smaller timestep (0.25 vs. 0.4~fs) is used. In the supplemental material to Ref.~\onlinecite{Mathew2018}, a study using a SB-type FF parameterized for TATB (2,4,6-triamino-1,3,5-trinitrobenzene), Mathew et al. mention a discontinuity of the potential energy function when the three-center angles approach 180$^\circ$. We wonder if a similar effect might be at play for RDX. The effect being very small ($<2\times10^{-9}$ eV/molecule/timestep), its impact could also have been overlooked, and in any case the issues was not investigated further and was deemed as irrelevant in the current study.

Armed with the heat flux determined from Eq.~\ref{eq:J}, and the thermal gradient extracted from the temperature profiles (Fig.~\ref{fig:RNEMD}), the thermal conductivity of the sample is then directly determined from Eq.~\ref{eq:fourier}. The last step is to take into account finite size effects that lead to scattering, as was mentioned in Ref.~\onlinecite{Schelling2002}. Therein, the authors discuss the specific task of simulating thermal conductivity in a solid under periodic boundary conditions, and show that the size effects lead to a linear dependence of the resistivity (1/$\kappa$) with respect to the inverse of the sample length $L_{\alpha}$:

\begin{equation}\label{eq:PBC}
\frac{1}{\kappa}=\frac{1}{\kappa_{\infty}}-\frac{\lambda}{L_{\alpha}},
\end{equation}

\noindent where $\kappa_{\infty}$ is the thermal conductivity extrapolated to an infinite sample (the value we report in this paper), and $\lambda$ a coefficient. For each case (orientation, T, P), we run simulations with samples of four different lengths, and the thermal conductivity is obtained by fitting these four data points with Eq.~\ref{eq:PBC}. Each simulation consists of 250000 timesteps of equilibration (100~ps), followed by 20000~swaps (5~ns), of which the last 10000 are used for production.

\subsection{MD sample preparation}
$\alpha$-RDX is an orthorhombic crystal with space group $Pbca$ and 8 molecules (168 atoms) per unit cell~\cite{McCrone1950, Choi1972}. In order to determine the equilibrium sample size for given P/T initial conditions, we first perform simulations with a system of 3$\times$3$\times$3 unit cells in the NPT (isothermal--isobaric) ensemble. Pressure and temperature are imposed via Nose-Hoover barostat/thermostat, with damping parameters of 100 and 1000 timesteps, respectively; the orthorhombic symmetry is imposed by fixing the three angles of the supercell at 90 degrees, but the three dimensions are allowed to vary independently. A timestep of 0.5~fs is used for the NPT simulations, and the lattice constants are averaged over the last 1~ns of a 2~ns run. In order to further validate the SB-FF, we also determined the lattice constants of $\gamma$-RDX (orthorhombic $Pca2_1$, 8 molecules/unit cell)~\cite{Davidson2008} as a function of pressure, using a 3$\times$3$\times$3 supercell.

For the thermal conductivity calculations, we built samples of 8, 12, 16, and 24 unit cells in length and a cross section set at 3$\times$3 unit cells. In order to investigate the anisotropy of the thermal conductivity, the directions along the three perpendicular lattice vectors of the crystal are considered  ($<$100$>$, $<$010$>$, and $<$001$>$). These specific crystal orientations were chosen since they produce an orthorhombic configuration that allows for PBC simulations without the use of specific techniques to build infinite samples in arbitrary directions (see Ref.~\onlinecite{Kroonblawd2016a}). The resulting samples contain between 12096 and 36288 atoms, with dimensions at 300~K and 0~GPa between 10.7 and 32.3~nm for samples oriented along $<$100$>$ (3.5$\times$3.2~nm$^2$ cross-section), 9.2 and 27.7~nm (4.0$\times$3.2~nm$^2$ cross-section) for $<$010$>$, and 8.4 and 25.3~nm for $<$001$>$ (4.0$\times$3.5~nm$^2$ cross-section). The size of the smallest longitudinal dimension is still near twice the estimated phonon mean-free path in $\beta$-HMX,~\cite{Chitsazi2020} which we use as rough estimate for RDX as well. 

Before running the $R$NEMD simulations, the samples with the appropriate lattice constant were equilibrated at the target temperature for 100~ps with a Langevin thermostat. The NVE runs between velocity swaps had a duration of 250~fs, in accordance with previous studies;~\cite{Kroonblawd2016} we also independently checked on shorter runs (5000 swaps of equilibration followed by 5000 swaps for production) that the NVE run duration did not impact the results, along with the cross-section size, see Table~\ref{tab:validation}. Apart from the smallest cross-section, thermal conductivities lie within $\sim$10\% of each other. The 3$\times$3 cross-section / 250~fs swap frequency combination was thus used in this work. A timestep of 0.4~fs is used for the $R$NEMD simulations.

\begin{table}
\caption{Sensitivity of the thermal conductivity $\kappa$ to the swap frequency and cross-section of the sample. System: $<$100$>$ $\alpha$-RDX, T=300~K, P=0~GPa, length=8 unit cells.} 
\begin{ruledtabular}
\begin{tabular}{c c c c c}
cross-section&swap frequency (fs)&$\kappa$ (W.m$^{-1}$.K$^{-1}$)\\
\hline
2$\times$2&250&0.409$\pm$0.010\\
3$\times$3&250&0.237$\pm$0.003\\
4$\times$4&250&0.261$\pm$0.004\\
3$\times$3&500&0.248$\pm$0.003\\
3$\times$3&750&0.232$\pm$0.002\\
\end{tabular}
\label{tab:validation}
\end{ruledtabular}
\end{table}

\subsection{Experimental}

Impulsively stimulated thermal scattering (ISTS) was performed on $<$100$>$, $<$010$>$, and $<$001$>$ oriented RDX crystals. RDX crystals were grown from acetone solution after removal of contaminants and imperfections from the starting material by recrystallization. RDX crystals were then indexed, cut, and polished for optical clarity for use in transmission in ISTS. The resulting slab geometry RDX crystals of approximately 7$\times$7$\times$0.7-1 mm$^3$ size were affixed to a rotation stage and oriented perpendicularly to the probe beam. A set of translation stages was used to position each crystal at the optimal probe volume for ISTS. 

\begin{figure}
\centering
\includegraphics[width=0.9\columnwidth]{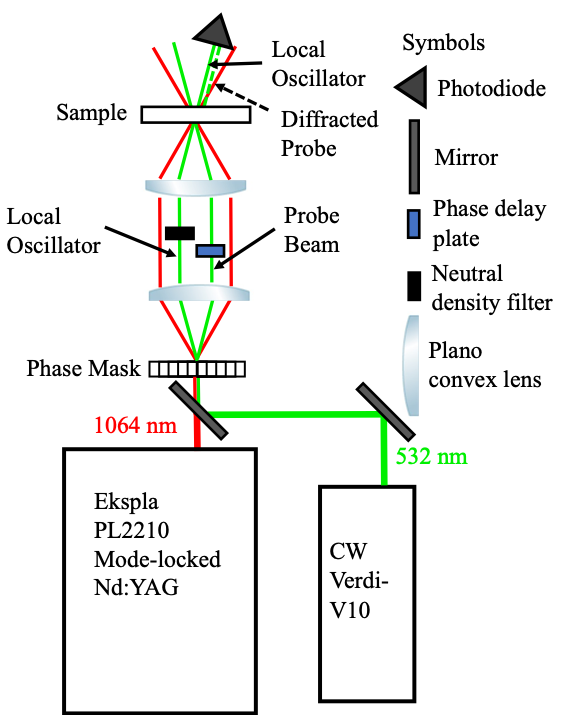}
\caption{\label{fig:exp}Schematic of the experimental apparatus. A box-CARS geometry was used to spatially separate the signal and pump beams after transmission through the sample. A 1.0 neutral density filter was placed in the local oscillator path. Due to the Bragg condition the local oscillator and diffracted probe are coincident after the sample.}
\end{figure}

ISTS is a non-contact optical technique for measuring thermal conductivity and sound speed in-situ.~\cite{1,2,3,4,5,6,7,8,9} A pair of 28 ps pump pulses (500 Hz repetition rate) at 1064 nm and continuous wave 532 probe beams were generated using the $\pm$1 order of a diffractive optical element. A spatial mask removed extraneous orders for both beams. The pump beams and probe were collimated and then focused into the sample. At the sample the pump and probe beams were approximately 100 $\mu$m and 50 $\mu$m in diameter respectively. Pump beams were crossed at $\sim$4.8$^\circ$ in the RDX sample. A small amount of the pump beam energy was absorbed by the crystal generating a thermally induced grating. The thermal excitation launched counter propagating pressure waves that travel at the sample acoustic velocity. The magnitude of the thermal grating decreases via conduction to ambient conditions. The probe beam was diffracted by the generated grating. The diffracted beam was sent to a DC-600 MHz silicon amplified detector (Thorlabs FPD610-FC-VIS) to measure the acoustic velocity and thermal grating decay. A 1~GHz oscilloscope was used to record the probed signal. Since RDX has weak absorption at the pump wavelength, the diffracted signal was small enough that a heterodyne technique was employed, wherein a spatially coincident local oscillator (LO) was mixed with the diffracted signal to simultaneously increase weak diffraction signals and reduce parasitic oscillator contributions.~\cite{5,6,7} A calcite phase plate was used to control the relative phase between the signal and LO beams. In phase and out of phase were defined to be 0$^\circ$ and 180$^\circ$, respectively, relative between the LO and signal beam phases. 512 shots were averaged per phase point during data collection. Thermal conductivity and sound speed data were collected at 5$^\circ$ radial increments through a full revolution. Shown in Fig.~\ref{fig:exp} is a schematic of the experimental apparatus. The LO and diffracted signal were spatially masked to reduce contributions from light scattering in the sample. Preliminary discussion of ISTS in $\alpha$-RDX was provided in Ref.~\onlinecite{LazarzSCCM}, although all results presented here were obtained from new measurements. 

\section{Results}
\subsection{Lattice constant and thermal expansion coefficient of $\alpha$-RDX as a function of pressure and temperature}\label{sec:lattice}

\begin{table*}
\centering
\caption{Comparison of the lattice constants and volume of the $\alpha$ and $\gamma$-RDX unit cells obtained in this work with the SB-FF to experimental and computational values from the literature. Error percentages with respect to experimental values are in parentheses.} 
\begin{ruledtabular}
\begin{tabular}{l | c c c c | c c c c c}
\multicolumn{5}{c}{$\alpha$-RDX at 300~K and 0~GPa}&\multicolumn{5}{c}{$\gamma$-RDX at 300~K and 5.2~GPa}\\
\hline
parameter&this work&DFT-D~[\onlinecite{Sorescu2010}]&DFT-D~[\onlinecite{Hunter2013}]&Exp~[\onlinecite{Olinger1978}]&&this work&SB-FF~[\onlinecite{Josyula2014}]&DFT-D~[\onlinecite{Sorescu2010}]&Exp~[\onlinecite{Davidson2008}]\\
a(\AA)&13.486 (+2.3)&13.237 (+0.4)&13.282 (+0.8)&13.182&&12.718 (+1.2)&12.71 (+1.2)&12.699 (+1.1)&12.565\\
b(\AA)&11.546 (-0.2)&11.391 (-1.6)&11.419 (-1.3)&11.574&&11.057 (+1.2)&11.05 (+1.1)&10.918 (-0.1)&10.930\\
c(\AA)&10.552 (-1.5)&10.770 (+0.6)&10.736 (+0.3)&10.709&&9.657 (+1.9)&9.64 (+1.7)&9.503 (+0.3)&9.477\\
V(\AA$^3$)&1642.93 (+0.6)&1623.94 (-0.6)&1628.27 (-0.3)&1633.86&&1357.91 (+4.3)&1354.9 (+4.1)&1317.66 (+1.2)&1301.5\\
\end{tabular}
\end{ruledtabular}
\label{tab:lattice}
\end{table*}

\begin{figure*}
\centering
\includegraphics[width=0.9\linewidth]{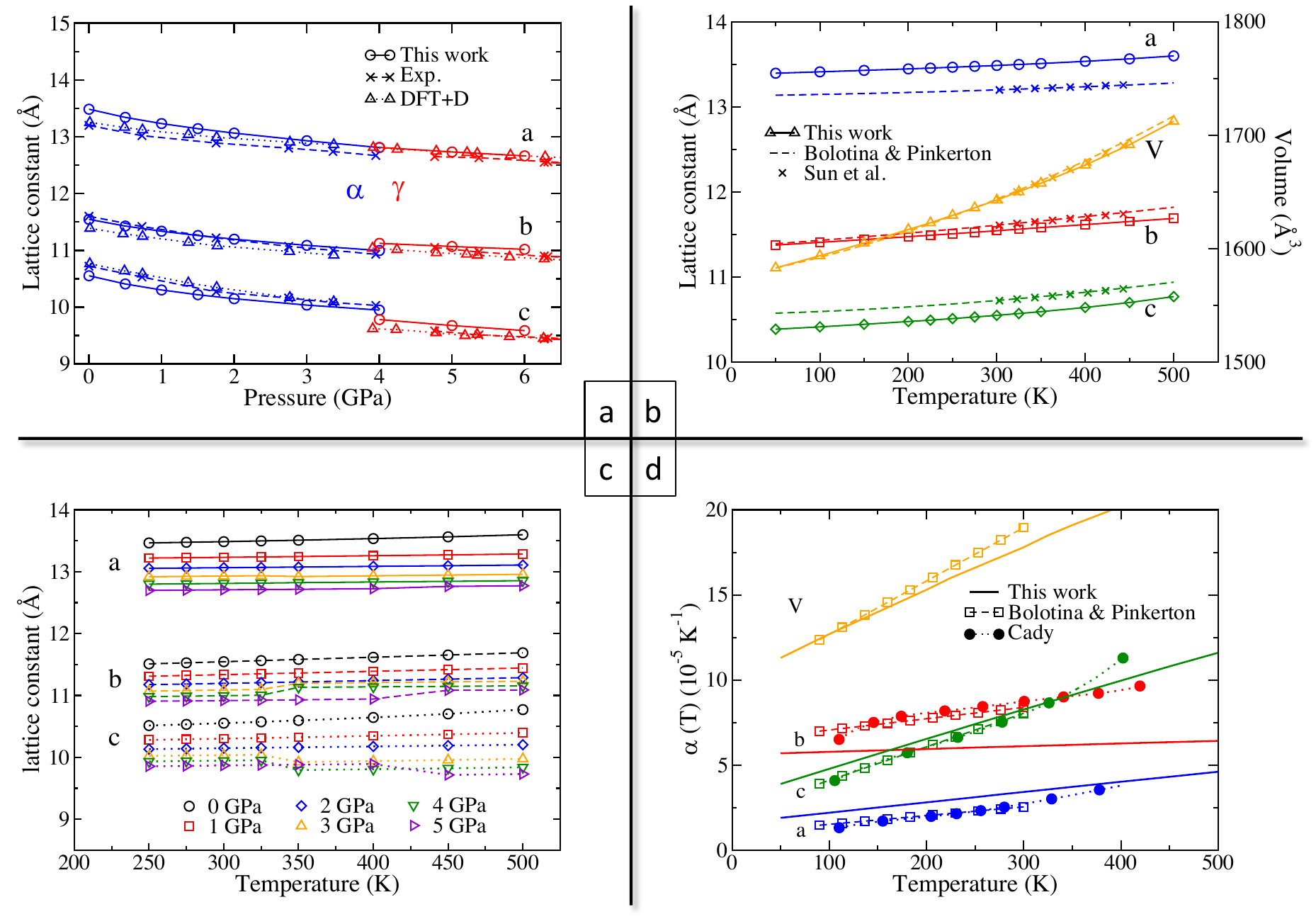}
\caption{\label{fig:valid}a: Lattice constants of $\alpha$ and $\gamma$-RDX as a function of pressure at 300~K, compared to experimental results from Refs.~\onlinecite{Olinger1978, Davidson2008} and zero temperature DFT+D results from Ref.~\onlinecite{Sorescu2010}. b: Lattice constants of $\alpha$-RDX as a function of temperature at 0~GPa from NPT-MD simulations, compared to experimental results from Bolotina and Pinkerton~\cite{Bolotina2015} and Sun et al..~\cite{Sun2011} c: Lattice constants of $\alpha$-RDX as a function of temperature and pressure from NPT-MD simulations. The cusps observed at P$\geq$3~GPa correspond to transition to the $\gamma$-phase during the MD simulations. d: Coefficient of thermal expansion in $\alpha$-RDX as a function of temperature, at 0~GPa. Results from this work are compared to experimental results from Cady~\cite{Cady1972} and Bolotina \& Pinkerton.~\cite{Bolotina2015}}
\end{figure*}

We first compare the equilibrium lattice constant and unit cell volume of $\alpha$ and $\gamma$-RDX (at 5.2~GPa for the latter), and compare it to experimental and $ab\ initio$ results, see Table~\ref{tab:lattice}. Included are results from DFT+D calculations (DFT with a correction for dispersion forces to reproduce crystalline equilibrium lattice constant). It can be seen that the SB-FF reproduces the lattice constants and unit cell volume with good accuracy, with an error compared to experiments of 0.6\% for the volume of the $\alpha$-RDX unit cell, and a maximum error on the lattice constant of 2.3\% (1.6\% for DFT+D). For $\gamma$-RDX, the error is below 5\% for the volume of the  unit cell (1.2\% for DFT+D), and 1.9\% at most for the lattice constants. Our results also agree well with previous calculations performed with the SB-FF for $\gamma$-RDX.

Figure~\ref{fig:valid}-a shows the lattice constants of $\alpha$ and $\gamma$-RDX as a function of pressure, at 300~K. Results obtained with the SB-FF (this work) are compared to experiments~\cite{Olinger1978, Davidson2008} and DFT+D calculations.~\cite{Sorescu2010} We again notice the good performance of the SB-FF, with errors on the order of the DFT+D errors. In addition, the qualitative changes in $a,b,c$ lattice constants between the $\alpha$ and $\gamma$ phases are reproduced correctly: $a$ shows almost no change, $b$ increases, $c$ decreases. Figure~\ref{fig:valid}-b reports the temperature dependence of the lattice constants and volume in $\alpha$-RDX, at zero pressure, and compared to experimental results using X-ray diffraction data.~\cite{Sun2011, Bolotina2015} The SB-FF reproduces qualitatively and quantitatively the experimental temperature dependence of the lattice constants and volume, with errors below 3\% for the lattice constant and excellent agreement on the volume, albeit through a compensation of errors on the $a$ and $c$ lattice constants. The increase between 50 and 500~K is 1.5\% for $a$ (1.1\% from experiments), 2.8\% for $b$ (3.8\%), and 3.7\% for $c$ (3.4\%). For the volume, our work predicts a total increase over the temperature range of 8.2\%, while Bolotina \& Pinkerton report 8.4\%.~\cite{Bolotina2015}

Lastly for the lattice constants, Fig.~\ref{fig:valid}-c reports the variation of $a$, $b$, and $c$ as a function of pressure between 250 and 500~K, in $\alpha$-RDX. We see that the pressure hinders the lattice expansion, with a marked effect on $c$ especially: the variation between the 250 and 500~K lattice constants goes from 1.0, 1.6, 2.5\% at 0~GPa to 0.4, 1.0, and 0.7\% at 2GPa, for $a$, $b$, and $c$ respectively. Additionally, we note a cusp for $b$ and $c$ above 325~K, starting at 3~GPa. This marks the phase transition to the $\gamma$ phase, with, accordingly, an increase in $b$ and a decrease in $c$, while $a$ remains about the same. While the experimental phase transition is reported closer to 4~GPa,~\cite{Davidson2008} the procedure employed here does not guarantee that 3~GPa is the ``true'' transition pressure with the SB-FF. Small systems used in MD simulations lead to artificial constraint through periodic boundary conditions, and the simulation timescale is orders of magnitude smaller than in experiments. What can be said with confidence is that simulations at 3~GPa and above should be closely monitored for a possible phase transition. 

\begin{table*}
\centering
\caption{Analytic expressions for the lattice constants and thermal expansion coefficients in $\alpha$-RDX as a function of pressure and temperature, compared to available experimental results. Lattice constants are given by L(T)=$a_0+a_1\times10^{-4}\times T+a_2\times10^{-7}\times T^2$, volume by V(T)=$a_0+a_1\times10^{-1}\times T+a_2\times10^{-4}\times T^2$, thermal expansion coefficients by $\alpha$(T)=$a_0\times10^{-5}+a_1\times10^{-7}\times T$. Last column denotes the temperature range over which the data was fitted.} 
\begin{ruledtabular}
\begin{tabular}{l | c c c | c c c | c c c | c c c | c c c | c c c c}
\multicolumn{14}{c}{Lattice constants and volume}\\
\hline
&\multicolumn{3}{c}{100}&\multicolumn{3}{c}{010}&\multicolumn{3}{c}{001}&\multicolumn{3}{c}{V}&T\\
&$a_0$&$a_1$&$a_2$&$a_0$&$a_1$&$a_2$&$a_0$&$a_1$&$a_2$&$a_0$&$a_1$&$a_2$\\
Ref.~\onlinecite{Bolotina2015}&13.1290&1.316&3.400&11.3557&7.236&4.163&10.5640&2.200&10.69&1575.14&1.463&2.745&90--300~K\\
P=0~GPa&13.3853&2.126&4.137&11.3440&6.363&1.146&10.3758&3.100&9.373&1575.75&1.563&2.280&50--500~K\\
P=1~GPa&13.1679&1.837&1.049&11.1796&5.147&0.302&10.1956&2.896&2.279&1501.08&1.308&0.569&250--500~K\\
P=2~GPa&13.0075&1.738&0.520&11.0792&3.490&1.315&10.0586&3.030&-0.246&1449.63&1.083&0.240&250--500~K\\
P=3~GPa&12.8803&1.046&0.200&10.9791&3.745&1.352&9.9506&3.296&-1.600&1407.16&1.061&-0.035&250--325~K\\
P=4~GPa&12.7500&1.894&0.400&10.9255&1.774&2.000&9.8428&4.597&-3.760&1371.09&1.0681&-0.021&250--325~K\\
P=5~GPa&12.6580&1.490&0.514&10.8760&0.544&2.819&9.7805&3.317&-1.291&1346.50&0.678&0.025&250--400~K\\
\end{tabular}
\begin{tabular}{l | c c  c | c  c c | c c c | c  c c | c c}
\multicolumn{14}{c}{Thermal expansion coefficients}\\
\hline
&\multicolumn{2}{c}{100}&&\multicolumn{2}{c}{010}&&\multicolumn{2}{c}{001}&&\multicolumn{2}{c}{V}&&T\\
&$a_0$&$a_1$&&$a_0$&$a_1$&&$a_0$&$a_1$&&$a_0$&$a_1$&\\
Ref.~\onlinecite{Bolotina2015}&1.007&0.512&&6.399&0.663&&2.134&1.97&&9.539&3.14&&90--300~K\\
Ref.~\onlinecite{Sun2011}&3.07&--&&8.28&--&&9.19&--&&20.7&--&&303--443~K\footnote{Fitted as constant}\\
Ref.~\onlinecite{Cady1972}&2.439&0.842&&8.531&0.673&&7.391&2.073&&18.33&3.625&&113--408~K\footnote{Terms in $T^2$ and beyond ignored, $<$100$>$ fitted up to 423~K.}\\
P=0~GPa&1.604&0.603&&5.618&0.163&&3.089&1.717&&10.30&2.478&&50--500~K\\
P=1~GPa&1.400&0.154&&4.608&0.031&&2.848&0.421&&8.854&0.606&&250--500~K\\
P=2~GPa&1.339&0.082&&3.167&0.225&&3.009&-0.056&&7.514&0.250&&250--500~K\\
P=3~GPa&0.818&0.307&&3.408&-0.082&&3.302&-0.325&&7.530&-0.101&&250--325~K\\
P=4~GPa&1.487&0.060&&1.634&-0.357&&4.641&-0.762&&7.761&-0.346&&250--325~K\\
P=5~GPa&1.179&0.079&&0.513&0.511&&3.380&-0.268&&5.067&0.323&&250--400~K\\
\end{tabular}
\end{ruledtabular}
\label{tab:CTEa}
\end{table*}

From these results, we extracted analytic functions to predict the lattice constants and volume as a function of pressure and temperature, which are reported in Table~\ref{tab:CTEa}. The quadratic expression also leads to a linear relation for the coefficient of linear thermal expansion in the direction $L$, $\alpha_L$(T):

\begin{equation}
\alpha_L(T)=\frac{1}{L}\frac{\textrm{d}L(T)}{\textrm{d}T},
\end{equation} 

\noindent and similarly for the coefficient of volume thermal expansion $\alpha_V$(T). The thermal expansion coefficients are also reported in Table~\ref{tab:CTEa}, and compare well with recent experimental results from Ref.~\onlinecite{Bolotina2015}. The thermal expansion coefficients are plotted in Fig.~\ref{fig:valid}-d, along with experimental results from Refs.~\onlinecite{Cady1972,Bolotina2015}.

\subsection{Thermal conductivity in $\alpha$-RDX as a function of temperature}\label{sec:T}

\begin{figure*}
\centering
\includegraphics[width=1\linewidth]{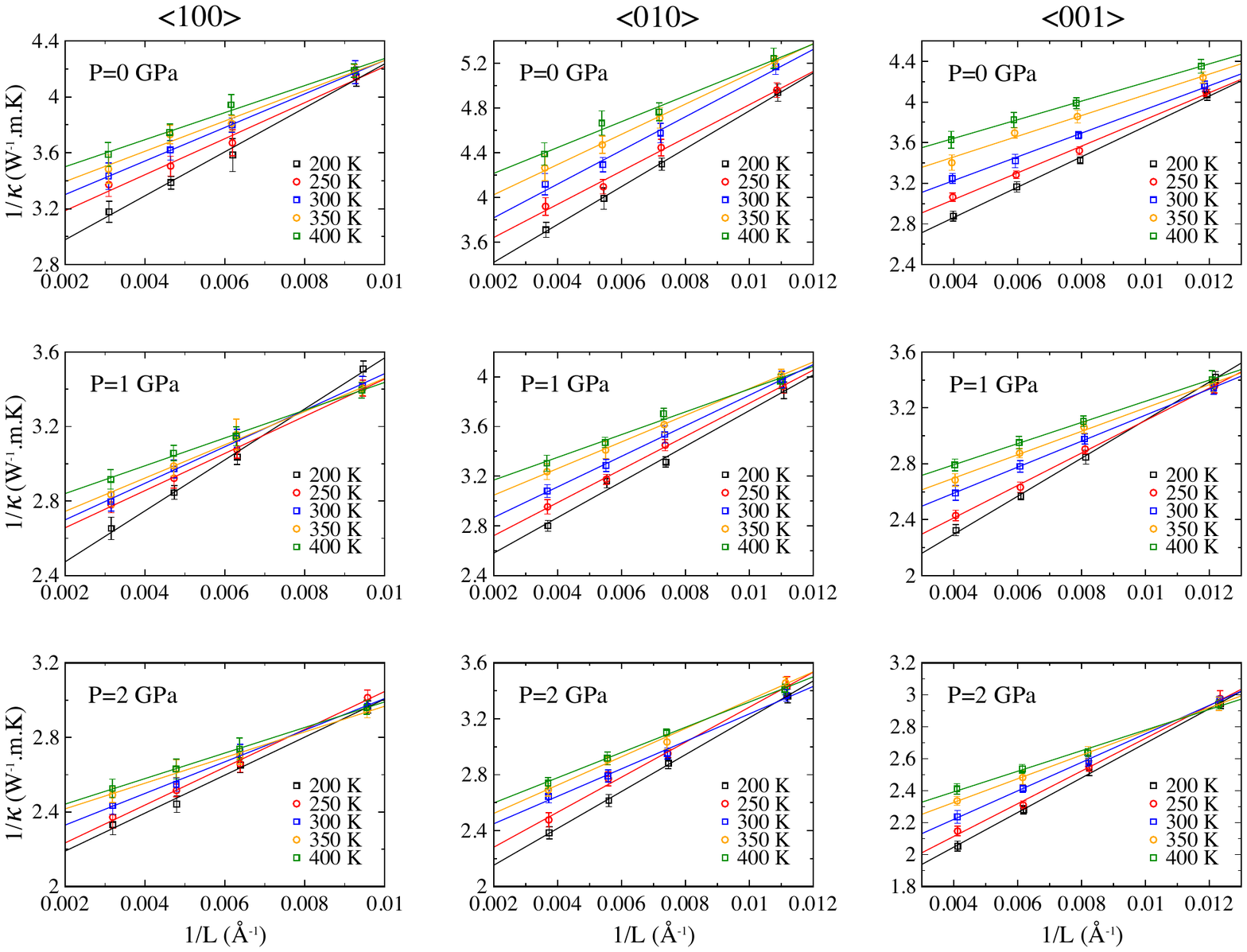}
\caption{\label{fig:kinf}Thermal resistivity (1/$\kappa$) as a function of the inverse sample length (1/$L$) and temperature for $\alpha$-RDX in the $<$100$>$, $<$010$>$, and $<$001$>$ orientations, at 0, 1, and 2~GPa. The solid lines are fit to the data using Eq.~\ref{eq:PBC}.}
\end{figure*}

In Fig.~\ref{fig:kinf}, we show the dependence of the thermal resistivity (1/$\kappa$) versus the inverse of the sample length (1/$L$), which is predicted to be linear according to Eq.~\ref{eq:PBC}. For each point, the thermal conductivity $\kappa(L,T,P)$ was obtained via Eq.~\ref{eq:fourier}. Heat exchanges were summed over 1000 swaps, and the sums averaged over 10000 swaps, which gives us the average heat flux $<$$\boldsymbol{J}$$>$ and the associated error $\delta(\boldsymbol{J})=\sigma_{\boldsymbol{J}}$ ($\sigma_{\boldsymbol{J}}$ is the standard deviation). The temperature gradient $\nabla(T)$ is obtained by fitting the temperature profiles, which produces two values since there are two directions of propagation (see Fig.~\ref{fig:RNEMD}); the standard deviation between the two values is used as error on $\nabla(T)$: $\delta(\nabla T)=\sigma_{\nabla T}$ (the standard error on the fit itself is negligible). Using standard error propagation rules~\cite{errorbook} and neglecting the covariance between $\nabla T$ and $\boldsymbol{J}$, the error $\delta(\kappa)$ on the thermal conductivity $\kappa$ is obtained as:

\begin{equation}
\delta(\kappa)=\sqrt{\kappa^2\left(\frac{\delta(\boldsymbol{J})^2}{<\boldsymbol{J}>^2}+\frac{\delta(\nabla T)^2}{<\nabla T>^2}\right)},
\end{equation}

\noindent and the error on the thermal resistivity $1/\kappa$ is then simply $\delta(1/\kappa)=\delta(\kappa)/\kappa^2$. All fits were performed with the R software package,~\cite{R} and each point was weighted by a factor equal to the inverse of the standard deviation of the measurement, when available, to account for the statistical uncertainty during fitting. 

\begin{figure*}
\centering
\includegraphics[width=0.9\linewidth]{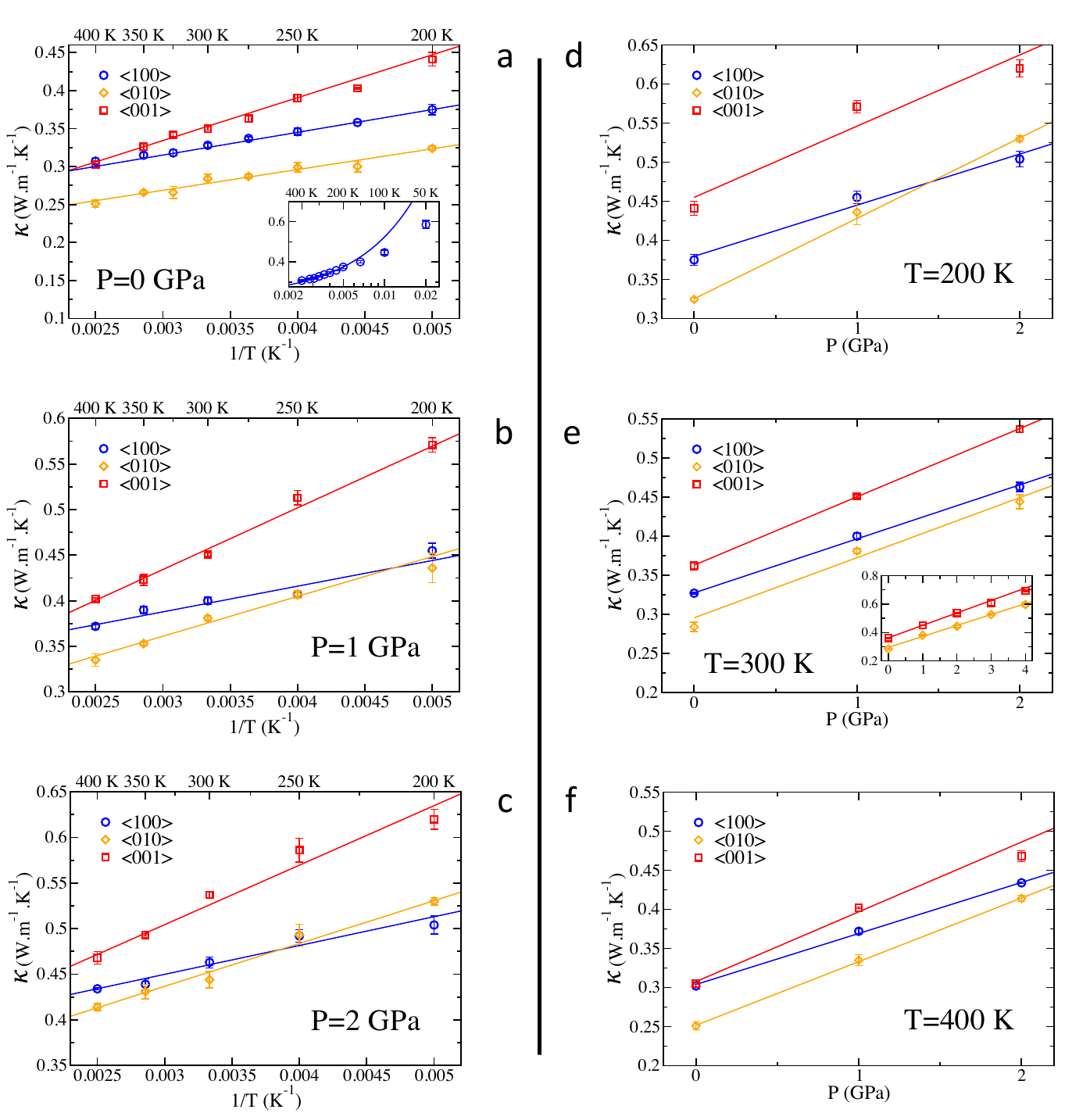}
\caption{\label{fig:TPdep}Thermal conductivity of $\alpha$-RDX in the $<$100$>$, $<$010$>$, and $<$001$>$ orientations, as a function of temperature at 0~GPa (a), 1~GPa (b), and 2~GPa (c), and as a function of pressure at 200~K (d), 300~K (e), and 400~K (f). Lines serve as guide to the eye (the fit to the data is discussed in Section~\ref{sec:final}). (a), insert: $<$100$>$ results down to 50~K, showing the breakdown of the linear behavior below 100--150~K. Note the log scale on the $x$ axis. (e), insert: $<$010$>$ and $<$001$>$ results up to 4~GPa, showing that the linear behavior holds at higher pressure.}
\end{figure*}

The thermal resistivity data in Fig.~\ref{fig:kinf} follows mostly a linear trend, which validates the behavior predicted by Eq.~\ref{eq:PBC}, and allows us to extract the extrapolated size-independent thermal conductivity, $\kappa_{\infty}$, which is the quantity most relevant to higher scale models and to compare to experiments. The results are reported in Figs.~\ref{fig:TPdep}-a--c for the temperature dependence. 

The first important information that can be obtained from Figs.~\ref{fig:TPdep}-a--c is that $\alpha$-RDX exhibits anisotropic thermal conductivity, with $\kappa_{001}>\kappa_{100}>\kappa_{010}$, and a difference of about 20\% between $\kappa_{001}$ and $\kappa_{010}$ at 300~K. Previous results obtained with the SB-FF and a slightly different $R$NEMD method by Izvekov et al. predict the same ordering for $\kappa_{001}$, $\kappa_{010}$, and $\kappa_{001}$, and a difference of 24.6\% between the two extremes at 300~K (more discussion in section~\ref{sec:discussion}).

It is also clear from Figs.~\ref{fig:TPdep}-a--c that the thermal conductivity of $\alpha$-RDX follows $\kappa\propto 1/T$ in the 200--400~K regime (simulations performed above 400~K occasionally led to a phase transition and were thus discarded). This is in agreement with theory that predicts a $1/T$ dependence  for a solid without defects at ``high'' temperature, i.e. $T>>T_{\Theta}$, where $T_{\Theta}$ is the Debye temperature of the solid.~\cite{Klemens1969} Using the isotropic continuum approximation and the sound speed extracted from the SB-FF and from experiments for $\alpha$-RDX, Ref.~\onlinecite{Izvekov2011} reports $T_{\Theta}$=100 and 125~K, respectively, while Rey-Lafon \& Bonjour estimate  $T_{\Theta}\simeq$60~K from the acoustic phonons at low temperature.~\cite{Izvekov2011,Reylafon1973} Standard theories show that the thermal conductivity varies as $\kappa\propto T^3$ at low temperature, however this is solely due to the temperature dependence of the heat capacity.~\cite{Kittel} With heat capacity being a constant in classical MD, there is no contribution to reduce the thermal conductivity at low temperature, and therefore a monotonic behavior can be expected. Additional calculations performed in the 50--200~K range for $\alpha$-RDX $<$100$>$ confirm a monotonous behavior, however the data departs from the linear behavior starting around 100--150~K (see Fig.~\ref{fig:TPdep}-a), suggesting that extrapolation of the trends extracted in the 200--400~K range to lower temperature should be taken with extreme caution. Interestingly, Izvekov et al.~\cite{Izvekov2011} reported a maximum of the thermal conductivity for single crystal RDX when using their NEMD method, at around 275~K, for all three directions. While that maximum seems to correlate with experimental observations compiled by Miller,~\cite{Miller1994} this is most likely fortuitous: Miller reports limited data points with large error bars, and used polycrystalline samples with density $\rho$=1.716~g/cm$^3$, vs. $\sim$1.8~g/cm$^3$ for ideal RDX crystals. In addition, a rigorous calculation with the Green-Kubo method, also performed in Ref.~\onlinecite{Izvekov2011}, did not reproduce this result, and the authors acknowledged the absence of a clear reason why a non-monotonic thermal conductivity was observed in a perfect crystal within the classical approximation.

\subsection{Thermal conductivity in $\alpha$-RDX as a function of pressure}\label{sec:P}

Figures~\ref{fig:TPdep}-d--f reports the thermal conductivity of $\alpha$-RDX between 0 and 2~GPa for each orientation, at $T$=200, 300, and 400~K (additional results from our calculations performed at 250 and 350~K are omitted for clarity). Preliminary calculations at 300~K showed that more than one simulation exhibited signs of a phase transition at 3 and 4~GPa, which prevented us from performing a reliable linear fit according to Eq.~\ref{eq:PBC}. This occurred specifically for samples in the $<$100$>$ orientation, but also for one case in the $<$001$>$ orientation, at 3~GPa (in that case the fit was still performed, based on 3 sample sizes). The onset of the phase transition is marked by changes in pressure and potential energy, however, since the volume of the sample is fixed at the beginning of the simulation and constrained throughout, the suspected transformation to $\gamma$ cannot be completed, and an unphysical structure is obtained. While the experimental $\alpha\rightarrow\gamma$ pressure phase transition is closer to 4~GPa, the SB-FF was shown to lead to phase transition at different pressures during uniaxial compression (depending on the axis), while transition under hydrostatic compression was not observed up to 9 ~GPa.~\cite{Munday2011} In addition to the accuracy of the force field itself, the size of the system and timescale of the simulation play a role here, since PBC introduce artificial constraints that can affect phase transition behavior. As a result, calculations were limited to the 0--2~GPa range, but the 3 and 4~GPa results at 300~K, are still presented in Fig.~\ref{fig:TPdep}-e to show that the trends still hold.

Our results suggest that $\kappa$ is a linear function of pressure. Specifically, $\kappa_{010}$ and  $\kappa_{001}$ follow similar quantitative trends, while  $\kappa_{100}$ is less affected by pressure. As a result, the difference between $\kappa_{010}$ and  $\kappa_{001}$ remains about the same as a function of pressure; $\kappa_{100}$ increases the most with respect to the other two directions, even crossing with $\kappa_{010}$ at 200~K, $\sim$1.5~GPa (the crossover pressure increases beyond 2~GPa at higher temperature, though the trends are similar). Experimentally, the linear trends should be measurable beyond uncertainties, with an increase of nearly 30\% between 0 and 2~GPa for $\kappa_{001}$ at 200~K, for instance.
 
While some theoretical models suggest a turnover of $\kappa(P)$ in molecular crystals at some pressure,~\cite{Long2012, Fan2017} this behavior is due to phonon softening, which usually occurs during a phase transition.~\cite{Nakanishi1982} Since our calculations were limited, by design, to a single phase of RDX, a monotonic relation between the thermal conductivity and pressure is to be expected. 

\subsection{Analytical expression for $\kappa(T,P)$}\label{sec:final}

Based on the results presented in sections~\ref{sec:T} and \ref{sec:P}, we propose a simple analytical expression that represents the $T$ and $P$-dependence of the thermal conductivity,

\begin{equation}
\kappa(T,P)=\kappa_0+\frac{\alpha}{T}+\beta P \label{eq:fit}
\end{equation}

\noindent where $a$,$b$, and $c$ are coefficients extracted, for each orientation, from the data obtained for $T$=200--400~K and $P$=0--2~GPa. The parameters are presented in Table~\ref{tab:fit}, and yield excellent agreement with the data: average errors between the predicted values and the data are less than 3\% for all orientations. A better agreement ($\leq$1.1\% on average for each orientation) can be obtained by using a more complex form, i.e. a second order function $\kappa(T,P)=a+b/T+c~P+d~P/T+f/T^2+g~P^2$; however, the underlying theory does not support a quadratic dependence on the inverse temperature, which may lead to drastic deviations when extrapolation is used. A different, empirical, form was recently employed in Ref.~\onlinecite{Leiding2020} to fit the average thermal conductivity in $\beta$-HMX, which incidentally seems to better match previous MD data for liquid HMX outside of the fitting range:  $\kappa=\kappa_0(\textrm{exp}[-\alpha T]+\beta P)$. Using this function did not lead to a significant improvement of the fit in our case and, as a result, the linear form of Eq.~\ref{eq:fit} is preferred. 

Finally, we used the parameters presented in Table~\ref{tab:fit} and Eq.~\ref{eq:fit} to map the thermal conductivity as a function of pressure and temperature, extrapolated to $T$=500~K and $P$=5~GPa, i.e. enclosing and exceeding the experimental domain of stability of $\alpha$-RDX.~\cite{Dreger2012} For each orientation, we observe that $\kappa(T,P)$ decreases monotonically from the high-$P$/low-$T$  region toward the low-$P$/high-$T$ region. 

\begin{table}
\centering
\setlength{\tabcolsep}{8pt}
\caption{Parameters obtained from fitting the $\kappa(T,P)$ data for each orientation, according to Eq.~\ref{eq:fit}. Fits performed to the data at T=200--400~K, P=0--2~GPa. Units are $\kappa_0$:~W.m$^{-1}$.K$^{-1}$; $\alpha$:~W.m$^{-1}$; $\beta$:~10$^{-9}$m$^2$.s$^{-1}$.K$^{-1}$ (=W.m$^{-1}$.K$^{-1}$.GPa$^{-1}$). Standard error on the parameters provided in parentheses.} 
\begin{tabular}{l c c c}
\hline
\hline
orientation&$\kappa_0$&$\alpha$&$\beta$\\
\hline
$<$100$>$&0.228 (0.007)&29.85 (1.99)&0.0667 (0.002)\\
$<$010$>$&0.156 (0.013)&36.75 (3.31)&0.0906 (0.004)\\
$<$001$>$&0.151 (0.013)&63.26 (3.98)&0.0841 (0.003)\\
\hline
\end{tabular}
\label{tab:fit}
\end{table}

\begin{figure}
\centering
\includegraphics[width=1\columnwidth]{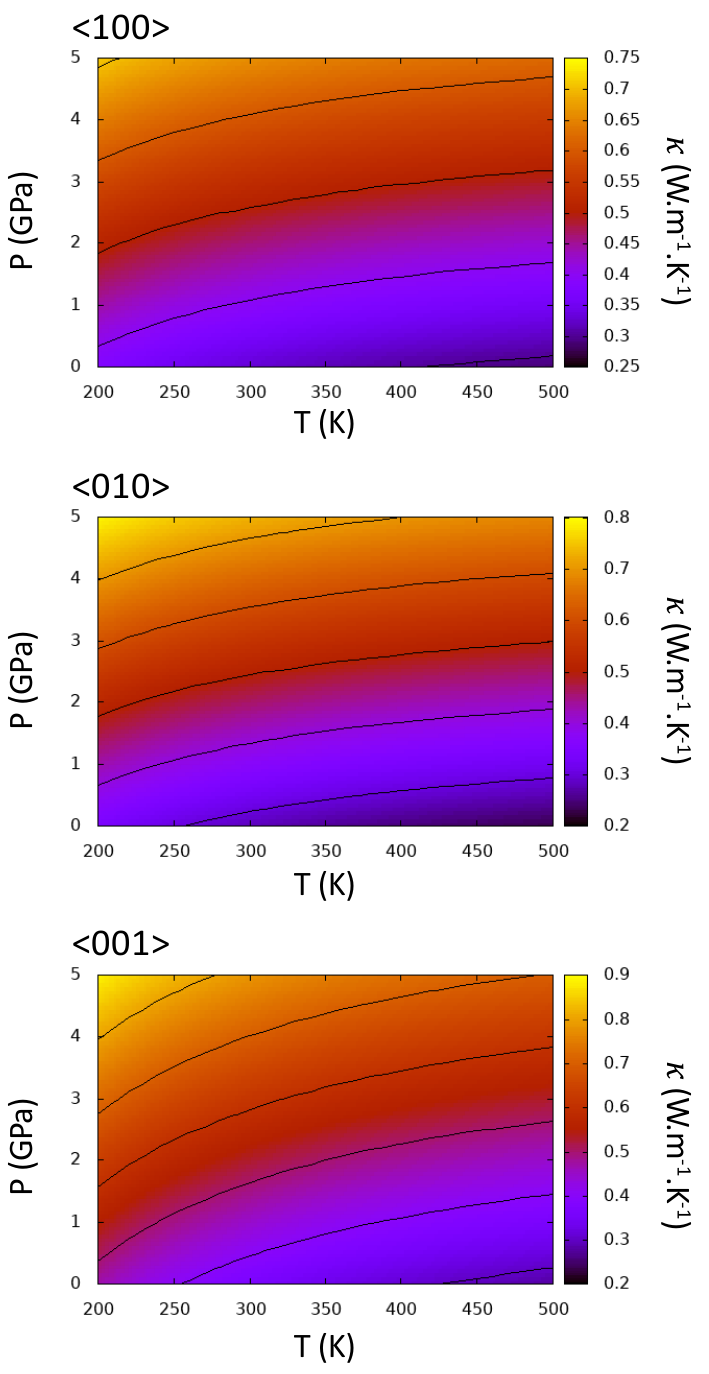}
\caption{\label{fig:final}Predicted thermal conductivity of $\alpha$-RDX in the $<$100$>$, $<$010$>$, and $<$001$>$ orientations, as a function of pressure and temperature up to 5~GPa and 500~K, using Eq.~\ref{eq:fit} and the parameters presented in Table~\ref{tab:fit}. Contours are plotted at 0.1~W.m$^{-1}$.K$^{-1}$ intervals.}
\end{figure}

\subsection{Experimental Results}\label{sec:ISTS}

\begin{figure*}
\centering
\includegraphics[width=0.9\linewidth]{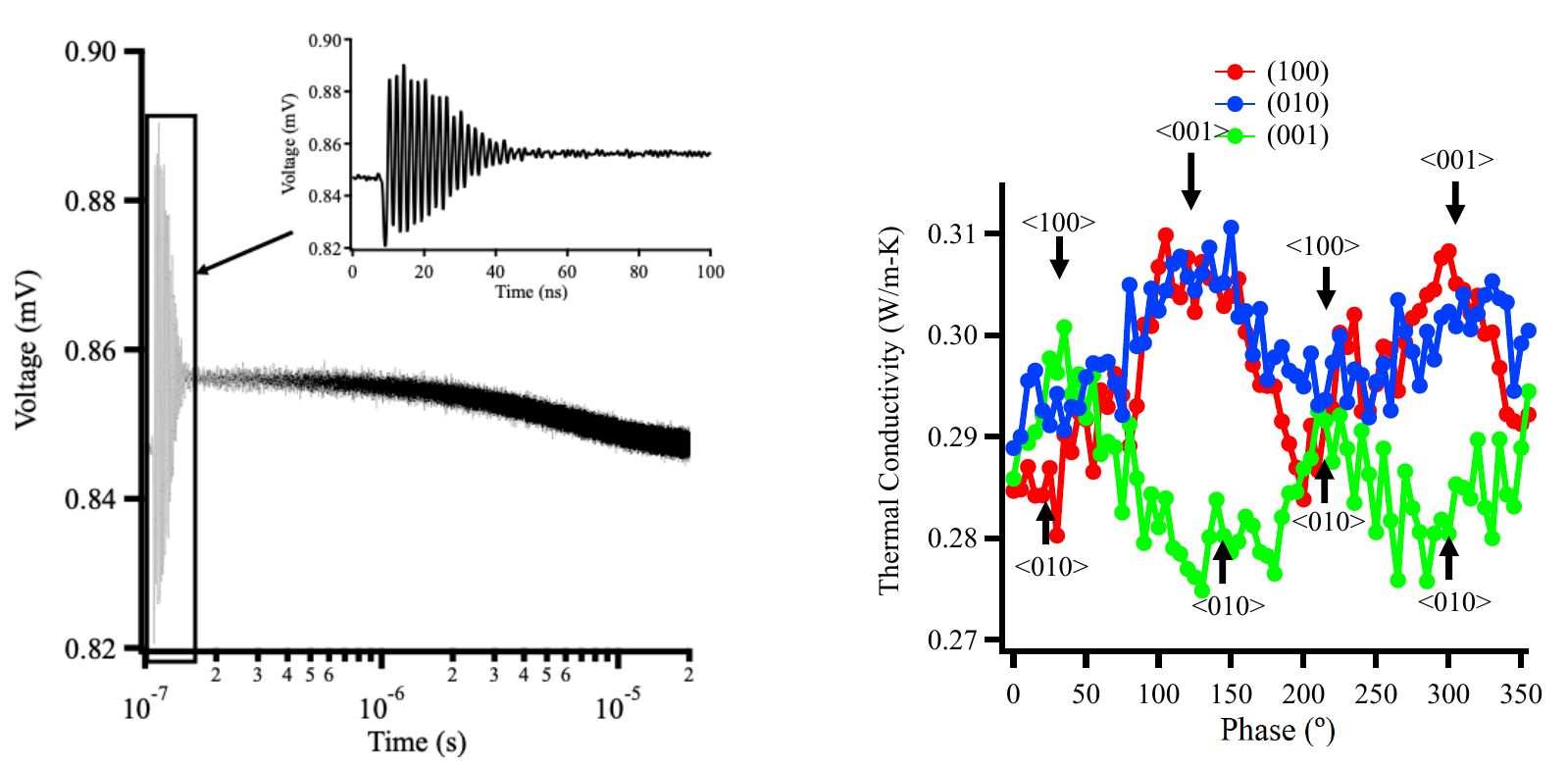}
\caption{\label{fig:Exp2}Left: Raw RDX $<$010$>$ ISTS measurements. The sound speed can be calculated from the frequency in the insert. The decay of the signal can be used to calculated the thermal conductivity. Right: Average thermal conductivities for $<$100$>$, $<$010$>$, and $<$001$>$ RDX crystal orientations through ISTS. Markers correspond to indexed crystal orientation, not the probed orientation. Probed directions are labelled at peaks and troughs.}
\end{figure*}

An RDX $<$010$>$ ISTS measurement is shown in Fig.~\ref{fig:Exp2}, left. Two time domain signals are present in the recorded measurement: the 10's nanosecond fast damped oscillatory portion and the ~20 microsecond slow decay. The damped oscillatory portion can be used to calculate the sound speed based on the frequency of the damped oscillations and the wavelength of the probe beam~\cite{6}. Rather than use a power spectrum fit~\cite{6,8,10} to find the frequency of the oscillatory portion, a nonlinear fit based on an analysis of variance was used,

\begin{equation}
I=\textrm{Acos}(f_{osc}t-B)e^{-Ct}+D
\end{equation}

where I is the measured signal, $A-D$ are fitting constants, and $f_{osc}$ is the frequency of the damped oscillations, and is directly related to the sound speed. The thermal conductivity can be calculated by the exponential decay of the signal after the acoustic portion. A moving average of 100 points was applied to both the fast and slow signals to decrease fitting times and further constrain the nonlinear fit to the measured signal.

Presented in Fig.~\ref{fig:Exp2}, right, are averaged experimental in phase and out of phase thermal conductivity results for RDX $<$100$>$, $<$010$>$, and $<$001$>$ oriented crystals. The reported thermal conductivities in Fig.~\ref{fig:Exp2}, right,  are for the indexed crystal orientation. The probed thermal conductivity was orthogonal to the reported orientation i.e. $<$100$>$ reported will result in probing of the $<$010$>$ and $<$001$>$ orientations. Probed orientations were inferred by comparing sound speed measurements performed in these experiments to velocity measurements from ISTS~\cite{10} and Brillouin scattering.~\cite{11} RDX has known anisotropy in sound speed,~\cite{10} and was expected to have anisotropy in thermal conductivity as well. The thermal conductivity oscillated between 0.28-0.31 W.m$^{-1}$.K$^{-1}$ for the $<$100$>$ RDX probed direction, 0.29-0.31 W.m$^{-1}$.K$^{-1}$ for $<$010$>$ RDX probed direction, and 0.275-0.30 W.m$^{-1}$.K$^{-1}$ for the $<$001$>$ RDX probed direction. The corresponding RDX orientations to probe direction are labelled in Fig.~\ref{fig:Exp2}, right. The average thermal conductivities are approximately 0.308, 0.298, and 0.283 W.m$^{-1}$.K$^{-1}$ along the $<$001$>$, $<$100$>$, and $<$010$>$  orientations, respectively.

The aperiodic noise present in the thermal conductivity measurements is mostly attributed to signal to detector alignment. Minor defects on the surface of crystals resulted in small pointing changes for the diffracted signal and LO into the photodiode. RDX crystal habits when formed from acetone lead to difficulty in cutting certain crystal facets, like the $<$001$>$ RDX orientation.~\cite{12, 13} The small pointing variations in the LO and signal resulted in smaller measured signals decreasing the signal to noise ratio. Additionally, the pointing variance increased deviation from the overall sinusoidal trends for the thermal conductivity as seen in Fig.~\ref{fig:Exp2}, left for the RDX $<$100$>$ probe direction. LO pointing was optimized on a shot to shot basis to increase the signal to noise ratio. 

\section{Discussion}\label{sec:discussion}

\begin{figure*}
\centering
\includegraphics[width=1.95\columnwidth]{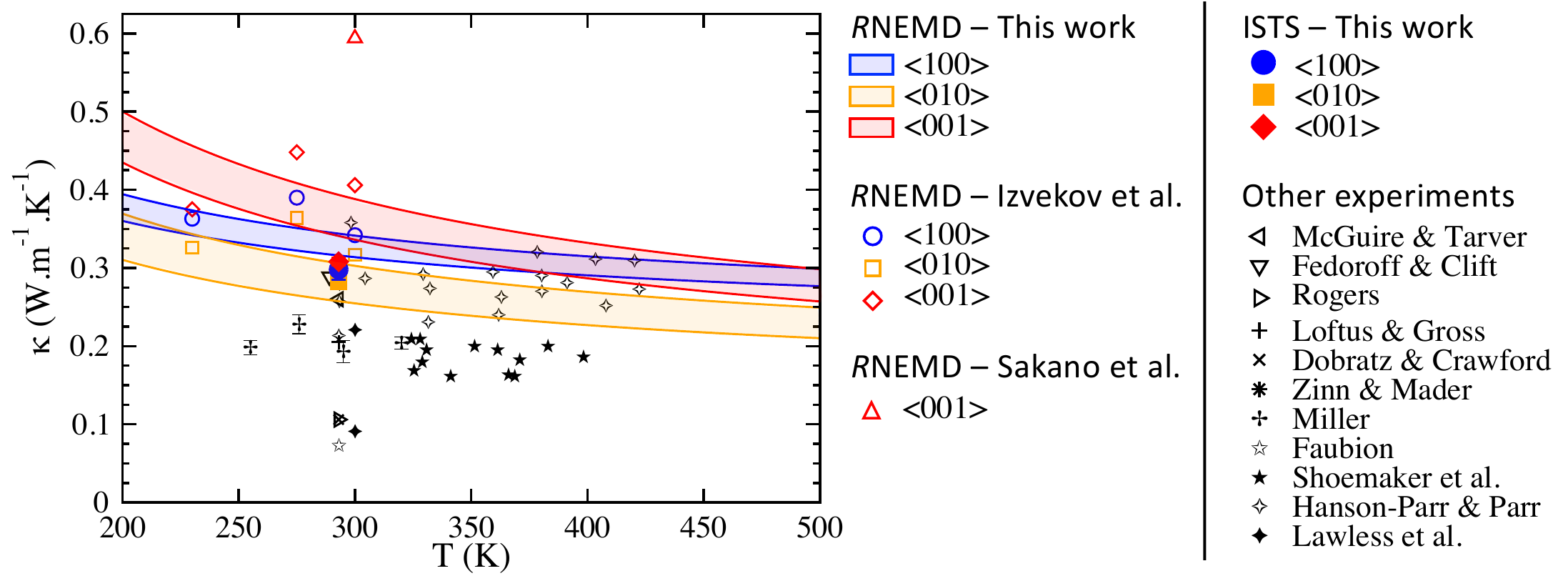}
\caption{\label{fig:Tdep}Comparison of the results from this work ($R$NEMD and ISTS) with $R$NEMD results from Izvekov et al.,~\cite{Izvekov2011} and Sakano et al.~\cite{Sakano2018}, and experimental results from McGuire \& Tarver,~\cite{McGuire1981} Fedoroff \& Clift,~\cite{Fedorov1960} Rogers,~\cite{Rogers1975} Loftus \& Gross,~\cite{Loftus1959} Dobratz \& Crawford,~\cite{Dobratz1985} Zinn \& Malder,~\cite{Zinn1960} Miller,~\cite{Miller1997} Faubion,~\cite{Faubion1976} Shoemaker et al.,~\cite{Shoemaker1985}, Hanson-Parr \& Parr,~\cite{Hanson-Parr1999} and Lawless et al..~\cite{Lawless2020} For the latter, the thermal conductivity is actually extracted from finite element simulations to reproduce experimental temperature profiles. The shaded region is bounded by extreme cases for $R$NEMD, determined from parameters and errors in Table~\ref{tab:fit}.}
\end{figure*}

\begin{table*}
\centering
\caption{Thermal conductivity of $\alpha$-RDX at near-ambient temperature and pressure conditions.} 
\begin{ruledtabular}
\begin{tabular}{l c c c c c c}
Source&&$\kappa$ (W.m$^{-1}$.K$^{-1}$)&T(K)&$\rho$ (g.cm$^3$)\\
\hline
This work $R$NEMD &$<$100$>$&0.316--0.344&293&1.80\\
&$<$010$>$&0.257--0.306&293&1.80\\
&$<$001$>$&0.341--0.394&293&1.80\\
This work ISTS &$<$100$>$&0.294--0.301&293&1.8\\
&$<$010$>$&0.280-0.287&293&1.8\\
&$<$001$>$&0.308-0.309&293&1.8\\
Izvekov et al.~\cite{Izvekov2011} ($R$NEMD)&$<$100$>$&0.342&300&1.805\\
&$<$010$>$&0.317&300&1.805\\
&$<$001$>$&0.406&300&1.805\\
Sakano et al.~\cite{Sakano2018} ($R$NEMD)&$<$100$>$&0.594&300&1.805\\
McGuire \& Tarver~\cite{McGuire1981}&-&0.260&293&1.8\\
Fedoroff \& Clift~\cite{Fedorov1960}&-&0.292&29.3&1.533\\
Rogers~\cite{Rogers1975}&-&0.106&293&1.806\\
Loftus \& Gross~\cite{Loftus1959}&-&0.205&293&1.650\\
Dobratz \& Crawford~\cite{Dobratz1985}&-&0.106&295&1.810\\
Zinn \& Malder~\cite{Zinn1960}&-&0.293&293&1.8\\
Miller~\cite{Miller1997}&-&0.193&293&1.716\\
Faubion~\cite{Faubion1976}&-&0.073&293&1.660\\
Hanson-Parr \& Parr~\cite{Hanson-Parr1999}&-&0.213--0.305\footnote{Authors applied an empirical correction to $\kappa$ in order to account for sample porosity.}&293&1.60--1.64\\
Lawless et al.~\cite{Lawless2020}&-&0.09--0.22&300&1.105--1.703\footnote{$\kappa$ is obtained via finite element analysis simulations to match experimental temperature profiles.}
\end{tabular}
\end{ruledtabular}
\label{tab:comp}
\end{table*}

The results above provide $\kappa(T,P)$ along the three primitive lattice vectors of $\alpha$-RDX, in a relevant regime of temperatures and pressures. There are other crystalline orientations that would be relevant in the crystal; for instance, $(210)$ and $(111)$ surfaces are observed in experimental samples,~\cite{Connick1969} and thermal conductivity through the direction perpendicular to those would be useful. However, $\alpha$-RDX has an orthorhombic unit cell (space group $Pbca$), where a two-rank tensor such as the thermal conductivity has only three non-zero components, all independent.~\cite{crystalbook} As a result, the results presented in this work allow for the full determination of the thermal conductivity, in any arbitrary direction.

We compare the thermal conductivity obtained from this work with results from the literature in Fig.~\ref{fig:Tdep} and Table~\ref{tab:comp}. We  use the parameters and associated errors provided in Table~\ref{tab:fit} to obtain limiting values for the thermal conductivity, which is bounded at $P$=0 by ($\kappa_0-\delta(\kappa_0))+\frac{\alpha-\delta(\alpha)}{T}$ and $(\kappa_0+\delta(\kappa_0))+\frac{\alpha+\delta(\alpha)}{T}$, see Fig.~\ref{fig:Tdep}. Firstly, we note the good agreement between our $R$NEMD results and the ISTS points obtained in this work (results from ISTS can resolve the anisotropy in thermal conductivity as the probe is diffracted only in the thermal grating direction): the trend for orientation dependence is the same for the two methods, and the values agree between 1 and 20\%, depending on the orientation, as it appears that the $R$NEMD overestimates $\kappa_{100}$ and $\kappa_{001}$. Taking into account the limiting cases shown in Fig.~\ref{fig:Tdep}, we find that $\kappa_{010}$ from ISTS is within the predicted range, while the values for $\kappa_{100}$ and $\kappa_{001}$ are now within 6 and 11\% of the lower predictions, respectively. Considering possible experimental uncertainty, the accuracy of the SB-FF, and additional errors arising from the simulations and fit of the data, this level of agreement is very encouraging. It also suggests that the results obtained with $R$NEMD can indeed be used to model the thermal conductivity of HE crystals for which experimental measurements are not available and non-trivial to obtain (pressure dependence, in particular, would require the precise alignment of crystals within diamond anvil cells and a simultaneous control of the temperature). Simulations can also guide experiments to focus on specific results, like the cross-over behavior observed in Fig.~\ref{fig:TPdep} between $\kappa_{100}$ and $\kappa_{010}$ as a function of pressure. 

As mentioned earlier, our $R$NEMD results also compare well with the work of Izvekov et al. using a similar method,~\cite{Izvekov2011} with $\kappa_{001}>\kappa_{100}>\kappa_{010}$ in the 225--300~K temperature range at zero pressure, and the average thermal conductivities $\bar{\kappa}=(\kappa_{100}+\kappa_{010}+\kappa_{001})/3$ at 300~K within 3\% of each other. Our results do not reproduce the maximum in the thermal conductivity observed by Izvekov et al. at 275~K, however that is most likely a fortuitous result, as discussed in Sec.~\ref{sec:T} and acknowledged in Ref.~\onlinecite{Izvekov2011}. An additional result obtained via $R$NEMD by Sakano et al.~\cite{Sakano2018} for $<$001$>$ oriented RDX is also shown, higher than the ISTS results by over 90\%, and both our ($\sim$60\%) and Izvekov's ($\sim$45\%) $R$NEMD results. Sakano et al. used a similar setup as ours, but different sample sizes and parameters, notably a swapping frequency of 3~ps vs. 250~fs in our case. While Sakano et al. state that smaller swap frequencies induce ``\it{nonconstant heat flux and too large thermal gradient}\normalfont '', we have not observed such issues. Similarly, while Sakano et al. fitted the $1/\kappa$ vs. $1/L$ relation for $L$ between 32 and 59~nm due to a breakdown of the linearity for smaller samples, ours and Izvekov's results show the appropriate linear behavior for samples between 8 and 25~nm (as seen in Fig.~\ref{fig:kinf}), and 6 to 55~nm, respectively. These results, in addition to previous considerations on sufficient sample size in a similar molecular crystal~\cite{Chitsazi2020}, and the good agreement with ISTS results, give us confidence that the present results are representative of $\alpha$-RDX. 

Finally, Fig.~\ref{fig:Tdep} shows the experimental results of Miller~\cite{Miller1997} and Shoemaker et al.,~\cite{Shoemaker1985} which were fitted to a non-monotonic relation by Miller, although at this scale the error bars make it difficult to assess the true form of the curve. Additional experimental results are shown, most of it for polycrystalline or powder samples with density lower than that of single crystal $\alpha$-RDX under similar conditions, with results averaged over many crystal orientations which lowered the thermal conductivity values. As such, direct comparison with our predictions for single crystals is difficult. However, our results agree well with the experimental results presented by Hanson-Parr \& Parr~\cite{Hanson-Parr1999} (see Fig.~\ref{fig:Tdep}), who also used powders but explicitly corrected for porosity. Thermal conductivity from different sources at or near ambient conditions are also summarized in Table~\ref{tab:comp}. Additional ISTS experiments are currently underway on oriented single crystals to probe more temperatures and test the $R$NEMD trends. 

\section{Conclusion}\label{sec:conclusions}

We have used $R$NEMD simulations with the Smith-Bharadwaj non-reactive force field to determine the thermal conductivity, $\kappa$, of $\alpha$-RDX in the $<$100$>$, $<$010$>$, and $<$001$>$ directions, as a function of temperature and pressure for $T$=200-400~K and $P$=0--2~GPa. The SB-FF is shown to reproduce the lattice constants and thermal expansion coefficients of $\alpha$-RDX in good agreement with experiments. We find that $\kappa$ varies linearly with the inverse temperature, in agreement with theory. In addition, we also observe a linear dependence of $\kappa$ with pressure. As a result, we suggest a simple linear form for $\kappa(T,P)$ for each orientation, and extract the corresponding coefficients which are then used to extrapolate $\kappa(T,P)$ up to 500~K and 5~GPa. 

Critically, ISTS measurements on RDX single crystals were performed to provide the first direct validation between MD and experimental results for orientation-dependent thermal conductivity in HE crystals. ISTS results validate the $R$NEMD anisotropy trend, and the numbers agree between 1 and 20\% depending on the orientation, for an average agreement around 10\%. These results can be used to parameterized mesoscale models that require anisotropic, temperature and pressure dependent, thermal properties for $\alpha$-RDX.

\section{Acknowledgements}
Work presented in this article was supported by the Laboratory Directed Research and Development program of Los Alamos National Laboratory under project number 20180100DR. This research used resources provided by the Los Alamos National Laboratory Institutional Computing Program. Los Alamos National Laboratory is operated by Triad National Security, LLC, for the National Nuclear Security Administration of U.S. Department of Energy (Contract No. 89233218CNA000001).

\bibliography{RDX}

\begin{thebibliography}{72}%
\makeatletter
\providecommand \@ifxundefined [1]{%
 \@ifx{#1\undefined}
}%
\providecommand \@ifnum [1]{%
 \ifnum #1\expandafter \@firstoftwo
 \else \expandafter \@secondoftwo
 \fi
}%
\providecommand \@ifx [1]{%
 \ifx #1\expandafter \@firstoftwo
 \else \expandafter \@secondoftwo
 \fi
}%
\providecommand \natexlab [1]{#1}%
\providecommand \enquote  [1]{``#1''}%
\providecommand \bibnamefont  [1]{#1}%
\providecommand \bibfnamefont [1]{#1}%
\providecommand \citenamefont [1]{#1}%
\providecommand \href@noop [0]{\@secondoftwo}%
\providecommand \href [0]{\begingroup \@sanitize@url \@href}%
\providecommand \@href[1]{\@@startlink{#1}\@@href}%
\providecommand \@@href[1]{\endgroup#1\@@endlink}%
\providecommand \@sanitize@url [0]{\catcode `\\12\catcode `\$12\catcode
  `\&12\catcode `\#12\catcode `\^12\catcode `\_12\catcode `\%12\relax}%
\providecommand \@@startlink[1]{}%
\providecommand \@@endlink[0]{}%
\providecommand \url  [0]{\begingroup\@sanitize@url \@url }%
\providecommand \@url [1]{\endgroup\@href {#1}{\urlprefix }}%
\providecommand \urlprefix  [0]{URL }%
\providecommand \Eprint [0]{\href }%
\providecommand \doibase [0]{http://dx.doi.org/}%
\providecommand \selectlanguage [0]{\@gobble}%
\providecommand \bibinfo  [0]{\@secondoftwo}%
\providecommand \bibfield  [0]{\@secondoftwo}%
\providecommand \translation [1]{[#1]}%
\providecommand \BibitemOpen [0]{}%
\providecommand \bibitemStop [0]{}%
\providecommand \bibitemNoStop [0]{.\EOS\space}%
\providecommand \EOS [0]{\spacefactor3000\relax}%
\providecommand \BibitemShut  [1]{\csname bibitem#1\endcsname}%
\let\auto@bib@innerbib\@empty
\bibitem [{\citenamefont {Bowden}\ and\ \citenamefont {Yoffe}(2009)}]{Bowden}%
  \BibitemOpen
  \bibfield  {author} {\bibinfo {author} {\bibfnamefont {F.~P.}\ \bibnamefont
  {Bowden}}\ and\ \bibinfo {author} {\bibfnamefont {Y.~D.}\ \bibnamefont
  {Yoffe}},\ }\href@noop {} {\emph {\bibinfo {title} {Initiation and Growth of
  Explosion in Liquids and Solids}}}\ (\bibinfo  {publisher} {Cambridge
  University Press},\ \bibinfo {year} {2009})\BibitemShut {NoStop}%
\bibitem [{\citenamefont {Storm}, \citenamefont {Stine},\ and\ \citenamefont
  {Kramer}(1990)}]{Storm1989}%
  \BibitemOpen
  \bibfield  {author} {\bibinfo {author} {\bibfnamefont {C.~B.}\ \bibnamefont
  {Storm}}, \bibinfo {author} {\bibfnamefont {J.~R.}\ \bibnamefont {Stine}}, \
  and\ \bibinfo {author} {\bibfnamefont {J.~F.}\ \bibnamefont {Kramer}},\
  }\bibfield  {title} {\enquote {\bibinfo {title} {Sensitivity relationships in
  energetic materials},}\ }in\ \href@noop {} {\emph {\bibinfo {booktitle}
  {{Chemistry and Physics of Energetic Materials}}}},\ \bibinfo {series} {NATO
  Advanced Science Institutes Series, Series C, Mathematical and Physical
  Sciences}, Vol.\ \bibinfo {volume} {309},\ \bibinfo {editor} {edited by\
  \bibinfo {editor} {\bibfnamefont {S.~N.}\ \bibnamefont {Bulusu}}}\ (\bibinfo
  {publisher} {Kluwer Academic Publ.},\ \bibinfo {year} {1990})\ pp.\ \bibinfo
  {pages} {605--639}\BibitemShut {NoStop}%
\bibitem [{\citenamefont {McGuire}\ and\ \citenamefont
  {Tarver}(1981)}]{McGuire1981}%
  \BibitemOpen
  \bibfield  {author} {\bibinfo {author} {\bibfnamefont {R.~R.}\ \bibnamefont
  {McGuire}}\ and\ \bibinfo {author} {\bibfnamefont {C.~M.}\ \bibnamefont
  {Tarver}},\ }\bibfield  {title} {\enquote {\bibinfo {title} {{Chemical
  decomposition models for the explosion of confined HMX, TATB, RDX, and TNT
  explosives}},}\ }in\ \href@noop {} {\emph {\bibinfo {booktitle} {Proceedings
  of the 7th International Symposium on Detonation}}}\ (\bibinfo  {publisher}
  {Office of Naval Research, Arlington, VA},\ \bibinfo {year} {1981})\ pp.\
  \bibinfo {pages} {56--64}\BibitemShut {NoStop}%
\bibitem [{\citenamefont {Fedoroff}\ and\ \citenamefont
  {Clift}(1960)}]{Fedorov1960}%
  \BibitemOpen
  \bibfield  {author} {\bibinfo {author} {\bibfnamefont {B.~T.}\ \bibnamefont
  {Fedoroff}}\ and\ \bibinfo {author} {\bibfnamefont {G.~D.}\ \bibnamefont
  {Clift}},\ }\href@noop {} {\emph {\bibinfo {title} {Encyclopedia of
  Explosives and Related Items}}}\ (\bibinfo  {publisher} {Picatinny Arsenal,
  Dover, NJ},\ \bibinfo {year} {1960})\BibitemShut {NoStop}%
\bibitem [{\citenamefont {Rogers}(1975)}]{Rogers1975}%
  \BibitemOpen
  \bibfield  {author} {\bibinfo {author} {\bibfnamefont {R.~N.}\ \bibnamefont
  {Rogers}},\ }\bibfield  {title} {\enquote {\bibinfo {title} {{Thermochemistry
  of explosives}},}\ }\href@noop {} {\bibfield  {journal} {\bibinfo  {journal}
  {Thermochim. Acta}\ }\textbf {\bibinfo {volume} {11}},\ \bibinfo {pages}
  {131--139} (\bibinfo {year} {1975})}\BibitemShut {NoStop}%
\bibitem [{\citenamefont {Loftus}\ and\ \citenamefont
  {Gross}(1959)}]{Loftus1959}%
  \BibitemOpen
  \bibfield  {author} {\bibinfo {author} {\bibfnamefont {G.}~\bibnamefont
  {Loftus}}\ and\ \bibinfo {author} {\bibfnamefont {G.}~\bibnamefont {Gross}},\
  }\href@noop {} {\emph {\bibinfo {title} {Thermal and Self Ignition Properties
  of Six Explosives, NBS Report 6548}}}\ (\bibinfo  {publisher} {National
  Bureau of Standards, Gaithersburg, MD},\ \bibinfo {year} {1959})\BibitemShut
  {NoStop}%
\bibitem [{\citenamefont {Dobratz}\ and\ \citenamefont
  {Crawford}(1985)}]{Dobratz1985}%
  \BibitemOpen
  \bibfield  {author} {\bibinfo {author} {\bibfnamefont {B.~M.}\ \bibnamefont
  {Dobratz}}\ and\ \bibinfo {author} {\bibfnamefont {P.~C.}\ \bibnamefont
  {Crawford}},\ }\href@noop {} {\emph {\bibinfo {title} {LLNL Explosives
  Handbook, Properties of Chemical Explosives and Explosive Simulants, Report
  UCRL 52997, Rev. 2}}}\ (\bibinfo  {publisher} {LLNL, Livermore, CA},\
  \bibinfo {year} {1985})\BibitemShut {NoStop}%
\bibitem [{\citenamefont {Zinn}\ and\ \citenamefont {Mader}(1960)}]{Zinn1960}%
  \BibitemOpen
  \bibfield  {author} {\bibinfo {author} {\bibfnamefont {J.}~\bibnamefont
  {Zinn}}\ and\ \bibinfo {author} {\bibfnamefont {C.~L.}\ \bibnamefont
  {Mader}},\ }\bibfield  {title} {\enquote {\bibinfo {title} {{Thermal
  initiation of explosives}},}\ }\href@noop {} {\bibfield  {journal} {\bibinfo
  {journal} {J. Appl. Phys.}\ }\textbf {\bibinfo {volume} {31}},\ \bibinfo
  {pages} {323--328} (\bibinfo {year} {1960})}\BibitemShut {NoStop}%
\bibitem [{\citenamefont {Miller}(1997)}]{Miller1997}%
  \BibitemOpen
  \bibfield  {author} {\bibinfo {author} {\bibfnamefont {M.~S.}\ \bibnamefont
  {Miller}},\ }\href@noop {} {\emph {\bibinfo {title} {Thermophysical
  Properties of RDX, ARL-TR-1319}}}\ (\bibinfo  {publisher} {Army Research Lab,
  Aberdeen Proving Ground, MD},\ \bibinfo {year} {1997})\BibitemShut {NoStop}%
\bibitem [{\citenamefont {Faubion}(1976)}]{Faubion1976}%
  \BibitemOpen
  \bibfield  {author} {\bibinfo {author} {\bibfnamefont {B.~D.}\ \bibnamefont
  {Faubion}},\ }\href@noop {} {\emph {\bibinfo {title} {Thermal Conductivity of
  RDX}}}\ (\bibinfo  {publisher} {Mason and Hanger-Silas Mason Co., Inc.,
  Amarillo, TX},\ \bibinfo {year} {1976})\BibitemShut {NoStop}%
\bibitem [{\citenamefont {Shoemaker}, \citenamefont {Stark},\ and\
  \citenamefont {Taylor}(1985)}]{Shoemaker1985}%
  \BibitemOpen
  \bibfield  {author} {\bibinfo {author} {\bibfnamefont {R.~L.}\ \bibnamefont
  {Shoemaker}}, \bibinfo {author} {\bibfnamefont {J.~A.}\ \bibnamefont
  {Stark}}, \ and\ \bibinfo {author} {\bibfnamefont {R.~E.}\ \bibnamefont
  {Taylor}},\ }\bibfield  {title} {\enquote {\bibinfo {title} {{Thermophysical
  properties of propellants}},}\ }\href@noop {} {\bibfield  {journal} {\bibinfo
   {journal} {HTHP}\ }\textbf {\bibinfo {volume} {17}},\ \bibinfo {pages}
  {429--435} (\bibinfo {year} {1985})}\BibitemShut {NoStop}%
\bibitem [{\citenamefont {Hanson-Parr}\ and\ \citenamefont
  {Parr}(1999)}]{Hanson-Parr1999}%
  \BibitemOpen
  \bibfield  {author} {\bibinfo {author} {\bibfnamefont {D.~M.}\ \bibnamefont
  {Hanson-Parr}}\ and\ \bibinfo {author} {\bibfnamefont {T.~P.}\ \bibnamefont
  {Parr}},\ }\bibfield  {title} {\enquote {\bibinfo {title} {{Thermal
  properties measurements of solid rocket propellant oxidizers and binder
  materials as a function of temperature}},}\ }\href@noop {} {\bibfield
  {journal} {\bibinfo  {journal} {J. Energ. Mater.}\ }\textbf {\bibinfo
  {volume} {17}},\ \bibinfo {pages} {1--48} (\bibinfo {year}
  {1999})}\BibitemShut {NoStop}%
\bibitem [{\citenamefont {Lawless}\ \emph {et~al.}(2020)\citenamefont
  {Lawless}, \citenamefont {Hobbs}, \citenamefont {Kaneshige}, \citenamefont
  {Lawless}, \citenamefont {Hobbs},\ and\ \citenamefont
  {Kaneshige}}]{Lawless2020}%
  \BibitemOpen
  \bibfield  {author} {\bibinfo {author} {\bibfnamefont {Z.~D.}\ \bibnamefont
  {Lawless}}, \bibinfo {author} {\bibfnamefont {M.~L.}\ \bibnamefont {Hobbs}},
  \bibinfo {author} {\bibfnamefont {M.~J.}\ \bibnamefont {Kaneshige}}, \bibinfo
  {author} {\bibfnamefont {Z.~D.}\ \bibnamefont {Lawless}}, \bibinfo {author}
  {\bibfnamefont {M.~L.}\ \bibnamefont {Hobbs}}, \ and\ \bibinfo {author}
  {\bibfnamefont {M.~J.}\ \bibnamefont {Kaneshige}},\ }\bibfield  {title}
  {\enquote {\bibinfo {title} {{Thermal conductivity of energetic
  materials}},}\ }\href@noop {} {\bibfield  {journal} {\bibinfo  {journal} {J.
  Energ. Mater.}\ }\textbf {\bibinfo {volume} {38}},\ \bibinfo {pages}
  {214--239} (\bibinfo {year} {2020})}\BibitemShut {NoStop}%
\bibitem [{\citenamefont {Kittel}(1966)}]{Kittel}%
  \BibitemOpen
  \bibfield  {author} {\bibinfo {author} {\bibfnamefont {C.}~\bibnamefont
  {Kittel}},\ }\bibfield  {title} {\enquote {\bibinfo {title} {Thermal
  poperties of insulators},}\ }in\ \href@noop {} {\emph {\bibinfo {booktitle}
  {Introduction to Solid State Physics, Third Edition}}}\ (\bibinfo
  {publisher} {John Wiley~\& Sons},\ \bibinfo {year} {1966})\ Chap.~\bibinfo
  {chapter} {6}\BibitemShut {NoStop}%
\bibitem [{\citenamefont {Kroonblawd}\ and\ \citenamefont
  {Sewell}(2016{\natexlab{a}})}]{Kroonblawd2016}%
  \BibitemOpen
  \bibfield  {author} {\bibinfo {author} {\bibfnamefont {M.~P.}\ \bibnamefont
  {Kroonblawd}}\ and\ \bibinfo {author} {\bibfnamefont {T.~D.}\ \bibnamefont
  {Sewell}},\ }\bibfield  {title} {\enquote {\bibinfo {title} {{Theoretical
  determination of anisotropic thermal conductivity for crystalline
  1,3,5-triamino-2,4,6-trinitrobenzene (TATB)}},}\ }\href@noop {} {\bibfield
  {journal} {\bibinfo  {journal} {J. Chem. Phys.}\ }\textbf {\bibinfo {volume}
  {139}},\ \bibinfo {pages} {074503} (\bibinfo {year}
  {2016}{\natexlab{a}})}\BibitemShut {NoStop}%
\bibitem [{\citenamefont {Kroonblawd}\ and\ \citenamefont
  {Sewell}(2016{\natexlab{b}})}]{Kroonblawd2016b}%
  \BibitemOpen
  \bibfield  {author} {\bibinfo {author} {\bibfnamefont {M.~P.}\ \bibnamefont
  {Kroonblawd}}\ and\ \bibinfo {author} {\bibfnamefont {T.~D.}\ \bibnamefont
  {Sewell}},\ }\bibfield  {title} {\enquote {\bibinfo {title} {{Theoretical
  determination of anisotropic thermal conductivity for initially defect-free
  and defective TATB single crystals}},}\ }\href@noop {} {\bibfield  {journal}
  {\bibinfo  {journal} {J. Chem. Phys.}\ }\textbf {\bibinfo {volume} {141}},\
  \bibinfo {pages} {184501} (\bibinfo {year} {2016}{\natexlab{b}})}\BibitemShut
  {NoStop}%
\bibitem [{\citenamefont {Bedrov}, \citenamefont {Smith},\ and\ \citenamefont
  {Sewell}(2000)}]{Bedrov2000}%
  \BibitemOpen
  \bibfield  {author} {\bibinfo {author} {\bibfnamefont {D.}~\bibnamefont
  {Bedrov}}, \bibinfo {author} {\bibfnamefont {G.~D.}\ \bibnamefont {Smith}}, \
  and\ \bibinfo {author} {\bibfnamefont {T.~D.}\ \bibnamefont {Sewell}},\
  }\bibfield  {title} {\enquote {\bibinfo {title} {{Thermal conductivity of
  liquid octahydro-1,3,5,7-tetranitro-1,3,5,7-tetrazocine (HMX) from molecular
  dynamics simulations}},}\ }\href@noop {} {\bibfield  {journal} {\bibinfo
  {journal} {Chem. Phys. Lett.}\ ,\ \bibinfo {pages} {64--68}} (\bibinfo {year}
  {2000})}\BibitemShut {NoStop}%
\bibitem [{\citenamefont {Chitsazi}\ \emph {et~al.}(2020)\citenamefont
  {Chitsazi}, \citenamefont {Kroonblawd}, \citenamefont {Pereverzev},\ and\
  \citenamefont {Sewell}}]{Chitsazi2020}%
  \BibitemOpen
  \bibfield  {author} {\bibinfo {author} {\bibfnamefont {R.}~\bibnamefont
  {Chitsazi}}, \bibinfo {author} {\bibfnamefont {M.~P.}\ \bibnamefont
  {Kroonblawd}}, \bibinfo {author} {\bibfnamefont {A.}~\bibnamefont
  {Pereverzev}}, \ and\ \bibinfo {author} {\bibfnamefont {T.}~\bibnamefont
  {Sewell}},\ }\bibfield  {title} {\enquote {\bibinfo {title} {{A molecular
  dynamics simulation study of thermal conductivity anisotropy in
  $\beta$-octahydro-1,3,5,7-tetranitro-1,3,5,7-tetrazocine ($\beta$-HMX) }},}\
  }\href@noop {} {\bibfield  {journal} {\bibinfo  {journal} {Model. Simul.
  Mater. Sci. Eng.}\ }\textbf {\bibinfo {volume} {28}},\ \bibinfo {pages}
  {025008} (\bibinfo {year} {2020})}\BibitemShut {NoStop}%
\bibitem [{\citenamefont {M{\"{u}}ller-Plathe}(1997)}]{Muller-Plathe1997}%
  \BibitemOpen
  \bibfield  {author} {\bibinfo {author} {\bibfnamefont {F.}~\bibnamefont
  {M{\"{u}}ller-Plathe}},\ }\bibfield  {title} {\enquote {\bibinfo {title} {{A
  simple nonequilibrium molecular dynamics method for calculating the thermal
  conductivity}},}\ }\href@noop {} {\bibfield  {journal} {\bibinfo  {journal}
  {J. Chem. Phys.}\ }\textbf {\bibinfo {volume} {106}},\ \bibinfo {pages}
  {6082} (\bibinfo {year} {1997})}\BibitemShut {NoStop}%
\bibitem [{\citenamefont {Zhang}\ \emph {et~al.}(2005)\citenamefont {Zhang},
  \citenamefont {Lussetti}, \citenamefont {de~Souza},\ and\ \citenamefont
  {M{\"{u}}ller-Plathe}}]{Zhang2005}%
  \BibitemOpen
  \bibfield  {author} {\bibinfo {author} {\bibfnamefont {M.}~\bibnamefont
  {Zhang}}, \bibinfo {author} {\bibfnamefont {E.}~\bibnamefont {Lussetti}},
  \bibinfo {author} {\bibfnamefont {L.~E.~S.}\ \bibnamefont {de~Souza}}, \ and\
  \bibinfo {author} {\bibfnamefont {F.}~\bibnamefont {M{\"{u}}ller-Plathe}},\
  }\bibfield  {title} {\enquote {\bibinfo {title} {{Thermal conductivities of
  molecular liquids by reverse nonequilibrium molecular dynamics}},}\
  }\href@noop {} {\bibfield  {journal} {\bibinfo  {journal} {J. Phys. Chem. B}\
  }\textbf {\bibinfo {volume} {109}},\ \bibinfo {pages} {15060--15067}
  (\bibinfo {year} {2005})}\BibitemShut {NoStop}%
\bibitem [{\citenamefont {Munday}\ \emph {et~al.}(2011)\citenamefont {Munday},
  \citenamefont {Chung}, \citenamefont {Rice},\ and\ \citenamefont
  {Solares}}]{Munday2011}%
  \BibitemOpen
  \bibfield  {author} {\bibinfo {author} {\bibfnamefont {L.~B.}\ \bibnamefont
  {Munday}}, \bibinfo {author} {\bibfnamefont {P.~W.}\ \bibnamefont {Chung}},
  \bibinfo {author} {\bibfnamefont {B.~M.}\ \bibnamefont {Rice}}, \ and\
  \bibinfo {author} {\bibfnamefont {S.~D.}\ \bibnamefont {Solares}},\
  }\bibfield  {title} {\enquote {\bibinfo {title} {{Simulations of
  high-pressure phases in RDX}},}\ }\href@noop {} {\bibfield  {journal}
  {\bibinfo  {journal} {J. Phys. Chem. B}\ }\textbf {\bibinfo {volume} {115}},\
  \bibinfo {pages} {4378--4386} (\bibinfo {year} {2011})}\BibitemShut {NoStop}%
\bibitem [{\citenamefont {Josyula}, \citenamefont {Rahul},\ and\ \citenamefont
  {De}(2014)}]{Josyula2014}%
  \BibitemOpen
  \bibfield  {author} {\bibinfo {author} {\bibfnamefont {K.}~\bibnamefont
  {Josyula}}, \bibinfo {author} {\bibnamefont {Rahul}}, \ and\ \bibinfo
  {author} {\bibfnamefont {S.}~\bibnamefont {De}},\ }\bibfield  {title}
  {\enquote {\bibinfo {title} {{Thermomechanical properties and equation of
  state for the gamma-polymorph of hexahydro-1,3,5-trinitro-1,3,5-triazine}},}\
  }\href@noop {} {\bibfield  {journal} {\bibinfo  {journal} {RSC Adv.}\
  }\textbf {\bibinfo {volume} {4}},\ \bibinfo {pages} {41491--41499} (\bibinfo
  {year} {2014})}\BibitemShut {NoStop}%
\bibitem [{\citenamefont {Hooks}\ \emph {et~al.}(2015)\citenamefont {Hooks},
  \citenamefont {Ramos}, \citenamefont {Bolme},\ and\ \citenamefont
  {Cawkwell}}]{Hooks2015}%
  \BibitemOpen
  \bibfield  {author} {\bibinfo {author} {\bibfnamefont {D.~E.}\ \bibnamefont
  {Hooks}}, \bibinfo {author} {\bibfnamefont {K.~J.}\ \bibnamefont {Ramos}},
  \bibinfo {author} {\bibfnamefont {C.~A.}\ \bibnamefont {Bolme}}, \ and\
  \bibinfo {author} {\bibfnamefont {M.~J.}\ \bibnamefont {Cawkwell}},\
  }\bibfield  {title} {\enquote {\bibinfo {title} {{Elasticity of crystalline
  molecular explosives}},}\ }\href@noop {} {\bibfield  {journal} {\bibinfo
  {journal} {Propellants, Explos. Pyrotech.}\ }\textbf {\bibinfo {volume}
  {40}},\ \bibinfo {pages} {333--350} (\bibinfo {year} {2015})}\BibitemShut
  {NoStop}%
\bibitem [{\citenamefont {Olinger}, \citenamefont {Roof},\ and\ \citenamefont
  {Cady}(1978)}]{Olinger1978}%
  \BibitemOpen
  \bibfield  {author} {\bibinfo {author} {\bibfnamefont {B.}~\bibnamefont
  {Olinger}}, \bibinfo {author} {\bibfnamefont {B.}~\bibnamefont {Roof}}, \
  and\ \bibinfo {author} {\bibfnamefont {H.}~\bibnamefont {Cady}},\ }\bibfield
  {title} {\enquote {\bibinfo {title} {The linear and volume compression of
  $\beta$-{HMX} and {RDX} to 9 {GPa}},}\ }in\ \href@noop {} {\emph {\bibinfo
  {booktitle} {Proc. Symposium (Intern.) on High Dynamic Pressures (C.E.A.,
  Paris, France)}}}\ (\bibinfo {year} {1978})\ pp.\ \bibinfo {pages}
  {3--8}\BibitemShut {NoStop}%
\bibitem [{\citenamefont {Davidson}\ \emph {et~al.}(2008)\citenamefont
  {Davidson}, \citenamefont {Oswald}, \citenamefont {Francis}, \citenamefont
  {Lennie}, \citenamefont {Marshall}, \citenamefont {Millar}, \citenamefont
  {Pulham}, \citenamefont {Warren},\ and\ \citenamefont
  {Cumming}}]{Davidson2008}%
  \BibitemOpen
  \bibfield  {author} {\bibinfo {author} {\bibfnamefont {A.~J.}\ \bibnamefont
  {Davidson}}, \bibinfo {author} {\bibfnamefont {I.~D.~H.}\ \bibnamefont
  {Oswald}}, \bibinfo {author} {\bibfnamefont {D.~J.}\ \bibnamefont {Francis}},
  \bibinfo {author} {\bibfnamefont {A.~R.}\ \bibnamefont {Lennie}}, \bibinfo
  {author} {\bibfnamefont {W.~G.}\ \bibnamefont {Marshall}}, \bibinfo {author}
  {\bibfnamefont {D.~I.~A.}\ \bibnamefont {Millar}}, \bibinfo {author}
  {\bibfnamefont {C.~R.}\ \bibnamefont {Pulham}}, \bibinfo {author}
  {\bibfnamefont {J.~E.}\ \bibnamefont {Warren}}, \ and\ \bibinfo {author}
  {\bibfnamefont {A.~S.}\ \bibnamefont {Cumming}},\ }\bibfield  {title}
  {\enquote {\bibinfo {title} {{Explosives under pressure--the crystal
  structure of $\gamma$-RDX as determined by high-pressure X-ray and neutron
  diffraction}},}\ }\href@noop {} {\bibfield  {journal} {\bibinfo  {journal}
  {Cryst. Eng. Comm.}\ }\textbf {\bibinfo {volume} {10}},\ \bibinfo {pages}
  {162--165} (\bibinfo {year} {2008})}\BibitemShut {NoStop}%
\bibitem [{\citenamefont {Dreger}(2012)}]{Dreger2012}%
  \BibitemOpen
  \bibfield  {author} {\bibinfo {author} {\bibfnamefont {Z.~A.}\ \bibnamefont
  {Dreger}},\ }\bibfield  {title} {\enquote {\bibinfo {title} {{Energetic
  materials under high pressures and temperatures: stability, polymorphism and
  decomposition of RDX}},}\ }\href@noop {} {\bibfield  {journal} {\bibinfo
  {journal} {J. Phys. Conf. Ser.}\ }\textbf {\bibinfo {volume} {377}},\
  \bibinfo {pages} {012047} (\bibinfo {year} {2012})}\BibitemShut {NoStop}%
\bibitem [{\citenamefont {Cawkwell}\ \emph {et~al.}(2016)\citenamefont
  {Cawkwell}, \citenamefont {Luscher}, \citenamefont {Addessio},\ and\
  \citenamefont {Ramos}}]{Cawkwell2016}%
  \BibitemOpen
  \bibfield  {author} {\bibinfo {author} {\bibfnamefont {M.~J.}\ \bibnamefont
  {Cawkwell}}, \bibinfo {author} {\bibfnamefont {D.~J.}\ \bibnamefont
  {Luscher}}, \bibinfo {author} {\bibfnamefont {F.~L.}\ \bibnamefont
  {Addessio}}, \ and\ \bibinfo {author} {\bibfnamefont {K.~J.}\ \bibnamefont
  {Ramos}},\ }\bibfield  {title} {\enquote {\bibinfo {title} {{Equations of
  state for the $\alpha$ and $\gamma$ polymorphs of cyclotrimethylene
  trinitramine}},}\ }\href@noop {} {\bibfield  {journal} {\bibinfo  {journal}
  {J. Appl. Phys.}\ }\textbf {\bibinfo {volume} {119}} (\bibinfo {year}
  {2016})}\BibitemShut {NoStop}%
\bibitem [{\citenamefont {Smith}\ and\ \citenamefont
  {Bharadwaj}(1999)}]{Smith1999}%
  \BibitemOpen
  \bibfield  {author} {\bibinfo {author} {\bibfnamefont {G.~D.}\ \bibnamefont
  {Smith}}\ and\ \bibinfo {author} {\bibfnamefont {R.~K.}\ \bibnamefont
  {Bharadwaj}},\ }\bibfield  {title} {\enquote {\bibinfo {title} {{Quantum
  chemistry based force field for simulations of HMX}},}\ }\href@noop {}
  {\bibfield  {journal} {\bibinfo  {journal} {J. Phys. Chem. B}\ }\textbf
  {\bibinfo {volume} {103}},\ \bibinfo {pages} {3570--3575} (\bibinfo {year}
  {1999})}\BibitemShut {NoStop}%
\bibitem [{\citenamefont {Bedrov}\ \emph {et~al.}(2001)\citenamefont {Bedrov},
  \citenamefont {Ayyagari}, \citenamefont {Smith}, \citenamefont {Sewell},
  \citenamefont {Menikoff},\ and\ \citenamefont {Zaug}}]{Bedrov2001}%
  \BibitemOpen
  \bibfield  {author} {\bibinfo {author} {\bibfnamefont {D.}~\bibnamefont
  {Bedrov}}, \bibinfo {author} {\bibfnamefont {C.}~\bibnamefont {Ayyagari}},
  \bibinfo {author} {\bibfnamefont {G.~D.}\ \bibnamefont {Smith}}, \bibinfo
  {author} {\bibfnamefont {T.~D.}\ \bibnamefont {Sewell}}, \bibinfo {author}
  {\bibfnamefont {R.}~\bibnamefont {Menikoff}}, \ and\ \bibinfo {author}
  {\bibfnamefont {J.~M.}\ \bibnamefont {Zaug}},\ }\bibfield  {title} {\enquote
  {\bibinfo {title} {{Molecular dynamics simulations of HMX crystal polymorphs
  using a flexible molecule force field}},}\ }\href@noop {} {\bibfield
  {journal} {\bibinfo  {journal} {J. Comput. Mater. Des.}\ }\textbf {\bibinfo
  {volume} {8}},\ \bibinfo {pages} {77--85} (\bibinfo {year}
  {2001})}\BibitemShut {NoStop}%
\bibitem [{\citenamefont {Weingarten}\ and\ \citenamefont
  {Sausa}(2015)}]{Weingarten2015}%
  \BibitemOpen
  \bibfield  {author} {\bibinfo {author} {\bibfnamefont {N.~S.}\ \bibnamefont
  {Weingarten}}\ and\ \bibinfo {author} {\bibfnamefont {R.~C.}\ \bibnamefont
  {Sausa}},\ }\bibfield  {title} {\enquote {\bibinfo {title} {{Nanomechanics of
  RDX single crystals by force displacement measurements and molecular dynamics
  simulations}},}\ }\href@noop {} {\bibfield  {journal} {\bibinfo  {journal}
  {J. Phys. Chem. A}\ }\textbf {\bibinfo {volume} {119}},\ \bibinfo {pages}
  {9338--9351} (\bibinfo {year} {2015})}\BibitemShut {NoStop}%
\bibitem [{\citenamefont {Josyula}, \citenamefont {Rahul},\ and\ \citenamefont
  {De}(2019)}]{Josyula2019}%
  \BibitemOpen
  \bibfield  {author} {\bibinfo {author} {\bibfnamefont {K.}~\bibnamefont
  {Josyula}}, \bibinfo {author} {\bibnamefont {Rahul}}, \ and\ \bibinfo
  {author} {\bibfnamefont {S.}~\bibnamefont {De}},\ }\bibfield  {title}
  {\enquote {\bibinfo {title} {{In silico study of $\alpha$-$\gamma$ phase
  transformation in hexahydro-1,3,5-trinitro-1,3,5-triazine}},}\ }\href@noop {}
  {\bibfield  {journal} {\bibinfo  {journal} {Comput. Mater. Sci.}\ }\textbf
  {\bibinfo {volume} {170}},\ \bibinfo {pages} {109180} (\bibinfo {year}
  {2019})}\BibitemShut {NoStop}%
\bibitem [{\citenamefont {Cawkwell}\ \emph {et~al.}(2010)\citenamefont
  {Cawkwell}, \citenamefont {Ramos}, \citenamefont {Hooks},\ and\ \citenamefont
  {Sewell}}]{Cawkwell2010}%
  \BibitemOpen
  \bibfield  {author} {\bibinfo {author} {\bibfnamefont {M.~J.}\ \bibnamefont
  {Cawkwell}}, \bibinfo {author} {\bibfnamefont {K.~J.}\ \bibnamefont {Ramos}},
  \bibinfo {author} {\bibfnamefont {D.~E.}\ \bibnamefont {Hooks}}, \ and\
  \bibinfo {author} {\bibfnamefont {T.~D.}\ \bibnamefont {Sewell}},\ }\bibfield
   {title} {\enquote {\bibinfo {title} {{Homogeneous dislocation nucleation in
  cyclotrimethylene trinitramine under shock loading}},}\ }\href@noop {}
  {\bibfield  {journal} {\bibinfo  {journal} {J. Appl. Phys.}\ }\textbf
  {\bibinfo {volume} {107}} (\bibinfo {year} {2010})}\BibitemShut {NoStop}%
\bibitem [{\citenamefont {Ramos}\ \emph {et~al.}(2010)\citenamefont {Ramos},
  \citenamefont {Hooks}, \citenamefont {Sewell},\ and\ \citenamefont
  {Cawkwell}}]{Ramos2010}%
  \BibitemOpen
  \bibfield  {author} {\bibinfo {author} {\bibfnamefont {K.~J.}\ \bibnamefont
  {Ramos}}, \bibinfo {author} {\bibfnamefont {D.~E.}\ \bibnamefont {Hooks}},
  \bibinfo {author} {\bibfnamefont {T.~D.}\ \bibnamefont {Sewell}}, \ and\
  \bibinfo {author} {\bibfnamefont {M.~J.}\ \bibnamefont {Cawkwell}},\
  }\bibfield  {title} {\enquote {\bibinfo {title} {{Anomalous hardening under
  shock compression in (021)-oriented cyclotrimethylene trinitramine single
  crystals}},}\ }\href@noop {} {\bibfield  {journal} {\bibinfo  {journal} {J.
  Appl. Phys.}\ }\textbf {\bibinfo {volume} {108}} (\bibinfo {year}
  {2010})}\BibitemShut {NoStop}%
\bibitem [{\citenamefont {Bidault}\ and\ \citenamefont
  {Pineau}(2018)}]{Bidault2018}%
  \BibitemOpen
  \bibfield  {author} {\bibinfo {author} {\bibfnamefont {X.}~\bibnamefont
  {Bidault}}\ and\ \bibinfo {author} {\bibfnamefont {N.}~\bibnamefont
  {Pineau}},\ }\bibfield  {title} {\enquote {\bibinfo {title} {{Granularity
  impact on hotspot formation and local chemistry in shocked nanostructured
  RDX}},}\ }\href@noop {} {\bibfield  {journal} {\bibinfo  {journal} {J. Chem.
  Phys.}\ }\textbf {\bibinfo {volume} {149}},\ \bibinfo {pages} {224703}
  (\bibinfo {year} {2018})}\BibitemShut {NoStop}%
\bibitem [{\citenamefont {Schelling}, \citenamefont {Phillpot},\ and\
  \citenamefont {Keblinski}(2002)}]{Schelling2002}%
  \BibitemOpen
  \bibfield  {author} {\bibinfo {author} {\bibfnamefont {P.~K.}\ \bibnamefont
  {Schelling}}, \bibinfo {author} {\bibfnamefont {S.~R.}\ \bibnamefont
  {Phillpot}}, \ and\ \bibinfo {author} {\bibfnamefont {P.}~\bibnamefont
  {Keblinski}},\ }\bibfield  {title} {\enquote {\bibinfo {title} {{Comparison
  of atomic-level simulation methods for computing thermal conductivity}},}\
  }\href@noop {} {\bibfield  {journal} {\bibinfo  {journal} {Phys. Rev. B}\
  }\textbf {\bibinfo {volume} {65}},\ \bibinfo {pages} {144306} (\bibinfo
  {year} {2002})}\BibitemShut {NoStop}%
\bibitem [{\citenamefont {Izvekov}, \citenamefont {Chung},\ and\ \citenamefont
  {Rice}(2011)}]{Izvekov2011}%
  \BibitemOpen
  \bibfield  {author} {\bibinfo {author} {\bibfnamefont {S.}~\bibnamefont
  {Izvekov}}, \bibinfo {author} {\bibfnamefont {P.~W.}\ \bibnamefont {Chung}},
  \ and\ \bibinfo {author} {\bibfnamefont {B.~M.}\ \bibnamefont {Rice}},\
  }\bibfield  {title} {\enquote {\bibinfo {title} {{Non-equilibrium molecular
  dynamics simulation study of heat transport in
  hexahydro-1,3,5-trinitro-s-triazine (RDX)}},}\ }\href@noop {} {\bibfield
  {journal} {\bibinfo  {journal} {Int. J. Heat Mass Transf.}\ }\textbf
  {\bibinfo {volume} {54}},\ \bibinfo {pages} {5623--5632} (\bibinfo {year}
  {2011})}\BibitemShut {NoStop}%
\bibitem [{\citenamefont {Plimpton}(1995)}]{Plimpton1995}%
  \BibitemOpen
  \bibfield  {author} {\bibinfo {author} {\bibfnamefont {S.}~\bibnamefont
  {Plimpton}},\ }\bibfield  {title} {\enquote {\bibinfo {title} {{Fast parallel
  algorithms for short-range molecular dynamics}},}\ }\href@noop {} {\bibfield
  {journal} {\bibinfo  {journal} {J. Comput. Phys.}\ }\textbf {\bibinfo
  {volume} {117}},\ \bibinfo {pages} {1--19} (\bibinfo {year}
  {1995})}\BibitemShut {NoStop}%
\bibitem [{\citenamefont {Mathew}\ \emph {et~al.}(2018)\citenamefont {Mathew},
  \citenamefont {Kroonblawd}, \citenamefont {Sewell},\ and\ \citenamefont
  {Thompson}}]{Mathew2018}%
  \BibitemOpen
  \bibfield  {author} {\bibinfo {author} {\bibfnamefont {N.}~\bibnamefont
  {Mathew}}, \bibinfo {author} {\bibfnamefont {M.~P.}\ \bibnamefont
  {Kroonblawd}}, \bibinfo {author} {\bibfnamefont {T.}~\bibnamefont {Sewell}},
  \ and\ \bibinfo {author} {\bibfnamefont {D.~L.}\ \bibnamefont {Thompson}},\
  }\bibfield  {title} {\enquote {\bibinfo {title} {{Predicted melt curve and
  liquid-state transport properties of TATB from molecular dynamics
  simulations}},}\ }\href@noop {} {\bibfield  {journal} {\bibinfo  {journal}
  {Mol. Simulat.}\ }\textbf {\bibinfo {volume} {44}},\ \bibinfo {pages}
  {613--622} (\bibinfo {year} {2018})}\BibitemShut {NoStop}%
\bibitem [{\citenamefont {McCrone}(1950)}]{McCrone1950}%
  \BibitemOpen
  \bibfield  {author} {\bibinfo {author} {\bibfnamefont {W.~C.}\ \bibnamefont
  {McCrone}},\ }\bibfield  {title} {\enquote {\bibinfo {title}
  {{Crystallographic data. 32. RDX (cyclotrimethylenetrinitramine)}},}\
  }\href@noop {} {\bibfield  {journal} {\bibinfo  {journal} {Anal. Chem.}\
  }\textbf {\bibinfo {volume} {22}},\ \bibinfo {pages} {954--955} (\bibinfo
  {year} {1950})}\BibitemShut {NoStop}%
\bibitem [{\citenamefont {Choi}\ and\ \citenamefont {Prince}(1972)}]{Choi1972}%
  \BibitemOpen
  \bibfield  {author} {\bibinfo {author} {\bibfnamefont {C.~S.}\ \bibnamefont
  {Choi}}\ and\ \bibinfo {author} {\bibfnamefont {E.}~\bibnamefont {Prince}},\
  }\bibfield  {title} {\enquote {\bibinfo {title} {The crystal structure of
  cyclotrimethylenetrinitramine},}\ }\href@noop {} {\bibfield  {journal}
  {\bibinfo  {journal} {Acta Crystallogr. B}\ }\textbf {\bibinfo {volume}
  {28}},\ \bibinfo {pages} {2857--2862} (\bibinfo {year} {1972})}\BibitemShut
  {NoStop}%
\bibitem [{\citenamefont {Kroonblawd}\ \emph {et~al.}(2016)\citenamefont
  {Kroonblawd}, \citenamefont {Mathew}, \citenamefont {Jiang},\ and\
  \citenamefont {Sewell}}]{Kroonblawd2016a}%
  \BibitemOpen
  \bibfield  {author} {\bibinfo {author} {\bibfnamefont {M.~P.}\ \bibnamefont
  {Kroonblawd}}, \bibinfo {author} {\bibfnamefont {N.}~\bibnamefont {Mathew}},
  \bibinfo {author} {\bibfnamefont {S.}~\bibnamefont {Jiang}}, \ and\ \bibinfo
  {author} {\bibfnamefont {T.~D.}\ \bibnamefont {Sewell}},\ }\bibfield  {title}
  {\enquote {\bibinfo {title} {{A generalized crystal-cutting method for
  modeling arbitrarily oriented crystals in 3D periodic simulation cells with
  applications to crystal–crystal interfaces}},}\ }\href@noop {} {\bibfield
  {journal} {\bibinfo  {journal} {Comput. Phys. Commun.}\ }\textbf {\bibinfo
  {volume} {207}},\ \bibinfo {pages} {232--242} (\bibinfo {year}
  {2016})}\BibitemShut {NoStop}%
\bibitem [{\citenamefont {Abramson}, \citenamefont {Brown},\ and\ \citenamefont
  {Slutsky}(1999)}]{1}%
  \BibitemOpen
  \bibfield  {author} {\bibinfo {author} {\bibfnamefont {E.~H.}\ \bibnamefont
  {Abramson}}, \bibinfo {author} {\bibfnamefont {J.~M.}\ \bibnamefont {Brown}},
  \ and\ \bibinfo {author} {\bibfnamefont {L.~J.}\ \bibnamefont {Slutsky}},\
  }\bibfield  {title} {\enquote {\bibinfo {title} {{Applications of impulsive
  stimulated scattering in the Earth and planetary sciences}},}\ }\href@noop {}
  {\bibfield  {journal} {\bibinfo  {journal} {Annu. Rev. Phys. Chem.}\ }\textbf
  {\bibinfo {volume} {50}},\ \bibinfo {pages} {279--313} (\bibinfo {year}
  {1999})}\BibitemShut {NoStop}%
\bibitem [{\citenamefont {Dennett}\ and\ \citenamefont {Short}(2018)}]{2}%
  \BibitemOpen
  \bibfield  {author} {\bibinfo {author} {\bibfnamefont {C.~A.}\ \bibnamefont
  {Dennett}}\ and\ \bibinfo {author} {\bibfnamefont {M.~P.}\ \bibnamefont
  {Short}},\ }\bibfield  {title} {\enquote {\bibinfo {title} {Thermal
  diffusivity determination using heterodyne phase insensitive transient
  grating spectroscopy},}\ }\href@noop {} {\bibfield  {journal} {\bibinfo
  {journal} {J. Appl. Phys.}\ }\textbf {\bibinfo {volume} {123}},\ \bibinfo
  {pages} {215109} (\bibinfo {year} {2018})}\BibitemShut {NoStop}%
\bibitem [{\citenamefont {Johnson}\ \emph {et~al.}(2012)\citenamefont
  {Johnson}, \citenamefont {Maznev}, \citenamefont {Bulsara}, \citenamefont
  {Fitzgerald}, \citenamefont {Harman}, \citenamefont {Calawa}, \citenamefont
  {Vineis}, \citenamefont {Turner},\ and\ \citenamefont {Nelson}}]{3}%
  \BibitemOpen
  \bibfield  {author} {\bibinfo {author} {\bibfnamefont {J.~A.}\ \bibnamefont
  {Johnson}}, \bibinfo {author} {\bibfnamefont {A.~A.}\ \bibnamefont {Maznev}},
  \bibinfo {author} {\bibfnamefont {M.~T.}\ \bibnamefont {Bulsara}}, \bibinfo
  {author} {\bibfnamefont {E.~A.}\ \bibnamefont {Fitzgerald}}, \bibinfo
  {author} {\bibfnamefont {T.~C.}\ \bibnamefont {Harman}}, \bibinfo {author}
  {\bibfnamefont {S.}~\bibnamefont {Calawa}}, \bibinfo {author} {\bibfnamefont
  {C.~J.}\ \bibnamefont {Vineis}}, \bibinfo {author} {\bibfnamefont
  {G.}~\bibnamefont {Turner}}, \ and\ \bibinfo {author} {\bibfnamefont {K.~A.}\
  \bibnamefont {Nelson}},\ }\bibfield  {title} {\enquote {\bibinfo {title}
  {Phase-controlled, heterodyne laser-induced transient grating measurements of
  thermal transport properties in opaque material},}\ }\href@noop {} {\bibfield
   {journal} {\bibinfo  {journal} {J. Appl. Phys.}\ }\textbf {\bibinfo {volume}
  {111}},\ \bibinfo {pages} {023503} (\bibinfo {year} {2012})}\BibitemShut
  {NoStop}%
\bibitem [{\citenamefont {K\"{a}ding}\ \emph {et~al.}(1995)\citenamefont
  {K\"{a}ding}, \citenamefont {Skurk}, \citenamefont {Maznev},\ and\
  \citenamefont {Matthias}}]{4}%
  \BibitemOpen
  \bibfield  {author} {\bibinfo {author} {\bibfnamefont {O.~W.}\ \bibnamefont
  {K\"{a}ding}}, \bibinfo {author} {\bibfnamefont {H.}~\bibnamefont {Skurk}},
  \bibinfo {author} {\bibfnamefont {A.~A.}\ \bibnamefont {Maznev}}, \ and\
  \bibinfo {author} {\bibfnamefont {E.}~\bibnamefont {Matthias}},\ }\bibfield
  {title} {\enquote {\bibinfo {title} {{Transient thermal gratings at surfaces
  for thermal characterization of bulk materials and thin films}},}\
  }\href@noop {} {\bibfield  {journal} {\bibinfo  {journal} {Appl. Phys. A}\
  }\textbf {\bibinfo {volume} {261}},\ \bibinfo {pages} {253--261} (\bibinfo
  {year} {1995})}\BibitemShut {NoStop}%
\bibitem [{\citenamefont {Maznev}, \citenamefont {Nelson},\ and\ \citenamefont
  {Rogers}(1998)}]{5}%
  \BibitemOpen
  \bibfield  {author} {\bibinfo {author} {\bibfnamefont {A.~A.}\ \bibnamefont
  {Maznev}}, \bibinfo {author} {\bibfnamefont {K.~A.}\ \bibnamefont {Nelson}},
  \ and\ \bibinfo {author} {\bibfnamefont {J.~A.}\ \bibnamefont {Rogers}},\
  }\bibfield  {title} {\enquote {\bibinfo {title} {Optical heterodyne detection
  of laser-induced gratings},}\ }\href@noop {} {\bibfield  {journal} {\bibinfo
  {journal} {Opt. Lett.}\ }\textbf {\bibinfo {volume} {23}},\ \bibinfo {pages}
  {1319--1321} (\bibinfo {year} {1998})}\BibitemShut {NoStop}%
\bibitem [{\citenamefont {Rogers}, \citenamefont {Maznev},\ and\ \citenamefont
  {Nelson}(2012)}]{6}%
  \BibitemOpen
  \bibfield  {author} {\bibinfo {author} {\bibfnamefont {J.~A.}\ \bibnamefont
  {Rogers}}, \bibinfo {author} {\bibfnamefont {A.}~\bibnamefont {Maznev}}, \
  and\ \bibinfo {author} {\bibfnamefont {K.~A.}\ \bibnamefont {Nelson}},\
  }\enquote {\bibinfo {title} {Impulsive stimulated thermal scattering},}\ in\
  \href@noop {} {\emph {\bibinfo {booktitle} {Characterization of Materials}}}\
  (\bibinfo  {publisher} {American Cancer Society},\ \bibinfo {year} {2012})\
  pp.\ \bibinfo {pages} {1--17}\BibitemShut {NoStop}%
\bibitem [{\citenamefont {Rogers}\ \emph {et~al.}(2000)\citenamefont {Rogers},
  \citenamefont {Maznev}, \citenamefont {Banet},\ and\ \citenamefont
  {Nelson}}]{7}%
  \BibitemOpen
  \bibfield  {author} {\bibinfo {author} {\bibfnamefont {J.~A.}\ \bibnamefont
  {Rogers}}, \bibinfo {author} {\bibfnamefont {A.~A.}\ \bibnamefont {Maznev}},
  \bibinfo {author} {\bibfnamefont {M.~J.}\ \bibnamefont {Banet}}, \ and\
  \bibinfo {author} {\bibfnamefont {K.~A.}\ \bibnamefont {Nelson}},\ }\bibfield
   {title} {\enquote {\bibinfo {title} {Optical generation and characterization
  of acoustic waves in thin films: Fundamentals and applications},}\
  }\href@noop {} {\bibfield  {journal} {\bibinfo  {journal} {Annu. Rev. Mater.
  Sci.}\ }\textbf {\bibinfo {volume} {30}},\ \bibinfo {pages} {117--157}
  (\bibinfo {year} {2000})}\BibitemShut {NoStop}%
\bibitem [{\citenamefont {Rogers}, \citenamefont {Yang},\ and\ \citenamefont
  {Nelson}(1994)}]{8}%
  \BibitemOpen
  \bibfield  {author} {\bibinfo {author} {\bibfnamefont {J.}~\bibnamefont
  {Rogers}}, \bibinfo {author} {\bibfnamefont {Y.}~\bibnamefont {Yang}}, \ and\
  \bibinfo {author} {\bibfnamefont {K.}~\bibnamefont {Nelson}},\ }\bibfield
  {title} {\enquote {\bibinfo {title} {Elastic modulus and in-plane thermal
  diffusivity measurements in thin polyimide films using symmetry-selective
  real-time impulsive stimulated thermal scattering},}\ }\href@noop {}
  {\bibfield  {journal} {\bibinfo  {journal} {Appl. Phys.}\ }\textbf {\bibinfo
  {volume} {58}},\ \bibinfo {pages} {523--534} (\bibinfo {year}
  {1994})}\BibitemShut {NoStop}%
\bibitem [{\citenamefont {Tokmakoff}, \citenamefont {Banholzer},\ and\
  \citenamefont {Fayer}(1993)}]{9}%
  \BibitemOpen
  \bibfield  {author} {\bibinfo {author} {\bibfnamefont {A.}~\bibnamefont
  {Tokmakoff}}, \bibinfo {author} {\bibfnamefont {W.~F.}\ \bibnamefont
  {Banholzer}}, \ and\ \bibinfo {author} {\bibfnamefont {M.~D.}\ \bibnamefont
  {Fayer}},\ }\bibfield  {title} {\enquote {\bibinfo {title} {Thermal
  diffusivity measurements of natural and isotopically enriched diamond by
  picosecond infrared transient grating experiments},}\ }\href@noop {}
  {\bibfield  {journal} {\bibinfo  {journal} {Appl. Phys. A}\ }\textbf
  {\bibinfo {volume} {56}},\ \bibinfo {pages} {87--90} (\bibinfo {year}
  {1993})}\BibitemShut {NoStop}%
\bibitem [{\citenamefont {Lazarz}\ \emph {et~al.}(2020)\citenamefont {Lazarz},
  \citenamefont {McGrane}, \citenamefont {Perriot}, \citenamefont {Bolme},
  \citenamefont {Cawkwell},\ and\ \citenamefont {Ramos}}]{LazarzSCCM}%
  \BibitemOpen
  \bibfield  {author} {\bibinfo {author} {\bibfnamefont {J.~D.}\ \bibnamefont
  {Lazarz}}, \bibinfo {author} {\bibfnamefont {S.~D.}\ \bibnamefont {McGrane}},
  \bibinfo {author} {\bibfnamefont {R.}~\bibnamefont {Perriot}}, \bibinfo
  {author} {\bibfnamefont {C.}~\bibnamefont {Bolme}}, \bibinfo {author}
  {\bibfnamefont {M.~J.}\ \bibnamefont {Cawkwell}}, \ and\ \bibinfo {author}
  {\bibfnamefont {K.~J.}\ \bibnamefont {Ramos}},\ }\bibfield  {title} {\enquote
  {\bibinfo {title} {{Anisotropic thermal conductivity and elasticity of RDX
  using impulsive stimulated thermal scattering}},}\ }\href@noop {} {\bibfield
  {journal} {\bibinfo  {journal} {AIP Conf. Proc.}\ }\textbf {\bibinfo {volume}
  {2272}},\ \bibinfo {pages} {060023} (\bibinfo {year} {2020})}\BibitemShut
  {NoStop}%
\bibitem [{\citenamefont {Sorescu}\ and\ \citenamefont
  {Rice}(2010)}]{Sorescu2010}%
  \BibitemOpen
  \bibfield  {author} {\bibinfo {author} {\bibfnamefont {D.~C.}\ \bibnamefont
  {Sorescu}}\ and\ \bibinfo {author} {\bibfnamefont {B.~M.}\ \bibnamefont
  {Rice}},\ }\bibfield  {title} {\enquote {\bibinfo {title} {{Theoretical
  predictions of energetic molecular crystals at ambient and hydrostatic
  compression conditions using dispersion corrections to conventional density
  functionals}},}\ }\href@noop {} {\bibfield  {journal} {\bibinfo  {journal}
  {J. Phys. Chem. C}\ }\textbf {\bibinfo {volume} {114}},\ \bibinfo {pages}
  {6734--6748} (\bibinfo {year} {2010})}\BibitemShut {NoStop}%
\bibitem [{\citenamefont {Hunter}\ \emph {et~al.}(2013)\citenamefont {Hunter},
  \citenamefont {Sutinen}, \citenamefont {Parker}, \citenamefont {Morrison},
  \citenamefont {Williamson}, \citenamefont {Thompson}, \citenamefont {Gould},\
  and\ \citenamefont {Pulham}}]{Hunter2013}%
  \BibitemOpen
  \bibfield  {author} {\bibinfo {author} {\bibfnamefont {S.}~\bibnamefont
  {Hunter}}, \bibinfo {author} {\bibfnamefont {T.}~\bibnamefont {Sutinen}},
  \bibinfo {author} {\bibfnamefont {S.~F.}\ \bibnamefont {Parker}}, \bibinfo
  {author} {\bibfnamefont {C.~A.}\ \bibnamefont {Morrison}}, \bibinfo {author}
  {\bibfnamefont {D.~M.}\ \bibnamefont {Williamson}}, \bibinfo {author}
  {\bibfnamefont {S.}~\bibnamefont {Thompson}}, \bibinfo {author}
  {\bibfnamefont {P.~J.}\ \bibnamefont {Gould}}, \ and\ \bibinfo {author}
  {\bibfnamefont {C.~R.}\ \bibnamefont {Pulham}},\ }\bibfield  {title}
  {\enquote {\bibinfo {title} {{Experimental and DFT‑D studies of the
  molecular organic energetic material RDX}},}\ }\href@noop {} {\bibfield
  {journal} {\bibinfo  {journal} {J. Phys. Chem. C}\ }\textbf {\bibinfo
  {volume} {117}},\ \bibinfo {pages} {8062--8071} (\bibinfo {year}
  {2013})}\BibitemShut {NoStop}%
\bibitem [{\citenamefont {Bolotina}\ and\ \citenamefont
  {Pinkerton}(2015)}]{Bolotina2015}%
  \BibitemOpen
  \bibfield  {author} {\bibinfo {author} {\bibfnamefont {N.~B.}\ \bibnamefont
  {Bolotina}}\ and\ \bibinfo {author} {\bibfnamefont {A.~A.}\ \bibnamefont
  {Pinkerton}},\ }\bibfield  {title} {\enquote {\bibinfo {title} {{Temperature
  dependence of thermal expansion tensors of energetic materials}},}\
  }\href@noop {} {\bibfield  {journal} {\bibinfo  {journal} {J. Appl.
  Crystallogr.}\ }\textbf {\bibinfo {volume} {48}},\ \bibinfo {pages}
  {1364--1380} (\bibinfo {year} {2015})}\BibitemShut {NoStop}%
\bibitem [{\citenamefont {Sun}\ \emph {et~al.}(2011)\citenamefont {Sun},
  \citenamefont {Shu}, \citenamefont {Liu}, \citenamefont {Zhang},
  \citenamefont {Liu}, \citenamefont {Jiang}, \citenamefont {Kang},
  \citenamefont {Xue},\ and\ \citenamefont {Song}}]{Sun2011}%
  \BibitemOpen
  \bibfield  {author} {\bibinfo {author} {\bibfnamefont {J.}~\bibnamefont
  {Sun}}, \bibinfo {author} {\bibfnamefont {X.}~\bibnamefont {Shu}}, \bibinfo
  {author} {\bibfnamefont {Y.}~\bibnamefont {Liu}}, \bibinfo {author}
  {\bibfnamefont {H.}~\bibnamefont {Zhang}}, \bibinfo {author} {\bibfnamefont
  {X.}~\bibnamefont {Liu}}, \bibinfo {author} {\bibfnamefont {Y.}~\bibnamefont
  {Jiang}}, \bibinfo {author} {\bibfnamefont {B.}~\bibnamefont {Kang}},
  \bibinfo {author} {\bibfnamefont {C.}~\bibnamefont {Xue}}, \ and\ \bibinfo
  {author} {\bibfnamefont {G.}~\bibnamefont {Song}},\ }\bibfield  {title}
  {\enquote {\bibinfo {title} {{Investigation on the thermal expansion and
  theoretical density of 1,3,5-trinitro-1,3,5-triazacyclohexane}},}\
  }\href@noop {} {\bibfield  {journal} {\bibinfo  {journal} {Propellants,
  Explos. Pyrotech.}\ }\textbf {\bibinfo {volume} {36}},\ \bibinfo {pages}
  {341--346} (\bibinfo {year} {2011})}\BibitemShut {NoStop}%
\bibitem [{\citenamefont {Cady}(1972)}]{Cady1972}%
  \BibitemOpen
  \bibfield  {author} {\bibinfo {author} {\bibfnamefont {H.~H.}\ \bibnamefont
  {Cady}},\ }\bibfield  {title} {\enquote {\bibinfo {title} {{Coefficient of
  thermal expansion of pentaerythritol tetranitrate and
  hexahydro-1,3,5-trinitro-1,3,5-triazine (RDX)}},}\ }\href@noop {} {\bibfield
  {journal} {\bibinfo  {journal} {J. Chem. Eng. Data}\ }\textbf {\bibinfo
  {volume} {17}},\ \bibinfo {pages} {369--371} (\bibinfo {year}
  {1972})}\BibitemShut {NoStop}%
\bibitem [{\citenamefont {Bevington}(1969)}]{errorbook}%
  \BibitemOpen
  \bibfield  {author} {\bibinfo {author} {\bibfnamefont {P.~R.}\ \bibnamefont
  {Bevington}},\ }in\ \href@noop {} {\emph {\bibinfo {booktitle} {Data
  Reduction and Error Analysis for the Physical Sciences}}}\ (\bibinfo
  {publisher} {McGraw-Hill Book Company},\ \bibinfo {year} {1969})\
  Chap.~\bibinfo {chapter} {4}\BibitemShut {NoStop}%
\bibitem [{\citenamefont {{R Core Team}}(2017)}]{R}%
  \BibitemOpen
  \bibfield  {author} {\bibinfo {author} {\bibnamefont {{R Core Team}}},\
  }\href {https://www.R-project.org/} {\emph {\bibinfo {title} {R: A Language
  and Environment for Statistical Computing}}},\ \bibinfo {organization} {R
  Foundation for Statistical Computing},\ \bibinfo {address} {Vienna, Austria}
  (\bibinfo {year} {2017})\BibitemShut {NoStop}%
\bibitem [{\citenamefont {Klemens}(1969)}]{Klemens1969}%
  \BibitemOpen
  \bibfield  {author} {\bibinfo {author} {\bibfnamefont {P.~G.}\ \bibnamefont
  {Klemens}},\ }\bibfield  {title} {\enquote {\bibinfo {title} {Thermal
  conductivity of solids},}\ }in\ \href@noop {} {\emph {\bibinfo {booktitle}
  {Thermal Conductivity}}},\ \bibinfo {editor} {edited by\ \bibinfo {editor}
  {\bibfnamefont {R.~P.}\ \bibnamefont {Tye}}}\ (\bibinfo  {publisher}
  {Academic Press},\ \bibinfo {year} {1969})\ Chap.~\bibinfo {chapter}
  {1}\BibitemShut {NoStop}%
\bibitem [{\citenamefont {Rey-Lafon}\ and\ \citenamefont
  {Bonjour}(1973)}]{Reylafon1973}%
  \BibitemOpen
  \bibfield  {author} {\bibinfo {author} {\bibfnamefont {M.}~\bibnamefont
  {Rey-Lafon}}\ and\ \bibinfo {author} {\bibfnamefont {E.}~\bibnamefont
  {Bonjour}},\ }\bibfield  {title} {\enquote {\bibinfo {title} {Étude de la
  chaleur specifique de la trinitro-1,3,5 hexahydro-s-triazine cristallisée
  détermination experimentale et calcul à partir des fréquences de vibration
  infrarouges et raman},}\ }\href@noop {} {\bibfield  {journal} {\bibinfo
  {journal} {Mol. Cryst. Liq. Cryst.}\ }\textbf {\bibinfo {volume} {24}},\
  \bibinfo {pages} {191--199} (\bibinfo {year} {1973})}\BibitemShut {NoStop}%
\bibitem [{\citenamefont {Miller}(1995)}]{Miller1994}%
  \BibitemOpen
  \bibfield  {author} {\bibinfo {author} {\bibfnamefont {M.~S.}\ \bibnamefont
  {Miller}},\ }\bibfield  {title} {\enquote {\bibinfo {title} {{Thermophysical
  properties of cyclotrimethylenetrinitramine}},}\ }\href@noop {} {\bibfield
  {journal} {\bibinfo  {journal} {J. Thermophys. Heat Transfer}\ }\textbf
  {\bibinfo {volume} {8}},\ \bibinfo {pages} {803--805} (\bibinfo {year}
  {1995})}\BibitemShut {NoStop}%
\bibitem [{\citenamefont {Long}\ \emph {et~al.}(2012)\citenamefont {Long},
  \citenamefont {Liu}, \citenamefont {Nie},\ and\ \citenamefont
  {Chen}}]{Long2012}%
  \BibitemOpen
  \bibfield  {author} {\bibinfo {author} {\bibfnamefont {Y.}~\bibnamefont
  {Long}}, \bibinfo {author} {\bibfnamefont {Y.~G.}\ \bibnamefont {Liu}},
  \bibinfo {author} {\bibfnamefont {F.-D.}\ \bibnamefont {Nie}}, \ and\
  \bibinfo {author} {\bibfnamefont {J.}~\bibnamefont {Chen}},\ }\bibfield
  {title} {\enquote {\bibinfo {title} {{A method to calculate the thermal
  conductivity of HMX under high pressure}},}\ }\href@noop {} {\bibfield
  {journal} {\bibinfo  {journal} {Philos. Mag.}\ }\textbf {\bibinfo {volume}
  {92}},\ \bibinfo {pages} {1023--1045} (\bibinfo {year} {2012})}\BibitemShut
  {NoStop}%
\bibitem [{\citenamefont {Fan}\ \emph {et~al.}(2017)\citenamefont {Fan},
  \citenamefont {Long}, \citenamefont {Ding}, \citenamefont {Chen},\ and\
  \citenamefont {Nie}}]{Fan2017}%
  \BibitemOpen
  \bibfield  {author} {\bibinfo {author} {\bibfnamefont {H.}~\bibnamefont
  {Fan}}, \bibinfo {author} {\bibfnamefont {Y.}~\bibnamefont {Long}}, \bibinfo
  {author} {\bibfnamefont {L.}~\bibnamefont {Ding}}, \bibinfo {author}
  {\bibfnamefont {J.}~\bibnamefont {Chen}}, \ and\ \bibinfo {author}
  {\bibfnamefont {F.-D.}\ \bibnamefont {Nie}},\ }\bibfield  {title} {\enquote
  {\bibinfo {title} {{A theoretical study of elastic anisotropy and thermal
  conductivity for TATB under pressure}},}\ }\href@noop {} {\bibfield
  {journal} {\bibinfo  {journal} {Comput. Mater. Sci.}\ }\textbf {\bibinfo
  {volume} {131}},\ \bibinfo {pages} {321--332} (\bibinfo {year}
  {2017})}\BibitemShut {NoStop}%
\bibitem [{\citenamefont {Nakanishi}, \citenamefont {Nagasawa},\ and\
  \citenamefont {Murakami}(1982)}]{Nakanishi1982}%
  \BibitemOpen
  \bibfield  {author} {\bibinfo {author} {\bibfnamefont {N.}~\bibnamefont
  {Nakanishi}}, \bibinfo {author} {\bibfnamefont {A.}~\bibnamefont {Nagasawa}},
  \ and\ \bibinfo {author} {\bibnamefont {Murakami}},\ }\bibfield  {title}
  {\enquote {\bibinfo {title} {{Lattice stability and soft modes}},}\
  }\href@noop {} {\bibfield  {journal} {\bibinfo  {journal} {J. Phys. Colloq.}\
  }\textbf {\bibinfo {volume} {43 (C4)}},\ \bibinfo {pages} {C4--35--C4--55}
  (\bibinfo {year} {1982})}\BibitemShut {NoStop}%
\bibitem [{\citenamefont {Leiding}\ \emph {et~al.}(2021)\citenamefont
  {Leiding}, \citenamefont {Velizhanin}, \citenamefont {Perriot}, \citenamefont
  {Cawkwell}, \citenamefont {Aslam},\ and\ \citenamefont
  {Andrews}}]{Leiding2020}%
  \BibitemOpen
  \bibfield  {author} {\bibinfo {author} {\bibfnamefont {J.~A.}\ \bibnamefont
  {Leiding}}, \bibinfo {author} {\bibfnamefont {K.~A.}\ \bibnamefont
  {Velizhanin}}, \bibinfo {author} {\bibfnamefont {R.}~\bibnamefont {Perriot}},
  \bibinfo {author} {\bibfnamefont {M.~J.}\ \bibnamefont {Cawkwell}}, \bibinfo
  {author} {\bibfnamefont {T.~D.}\ \bibnamefont {Aslam}}, \ and\ \bibinfo
  {author} {\bibfnamefont {S.~A.}\ \bibnamefont {Andrews}},\ }\bibfield
  {title} {\enquote {\bibinfo {title} {{A new parameterization of a 1-step
  thermal decomposition model of PBX-9501}},}\ }\href@noop {} {\bibfield
  {journal} {\bibinfo  {journal} {to be sumitted}\ } (\bibinfo {year}
  {2021})}\BibitemShut {NoStop}%
\bibitem [{\citenamefont {Sun}\ \emph {et~al.}(2008)\citenamefont {Sun},
  \citenamefont {Winey}, \citenamefont {Hemmi}, \citenamefont {Dreger},
  \citenamefont {Zimmerman}, \citenamefont {Gupta}, \citenamefont
  {Torchinsky},\ and\ \citenamefont {Nelson}}]{10}%
  \BibitemOpen
  \bibfield  {author} {\bibinfo {author} {\bibfnamefont {B.}~\bibnamefont
  {Sun}}, \bibinfo {author} {\bibfnamefont {J.~M.}\ \bibnamefont {Winey}},
  \bibinfo {author} {\bibfnamefont {N.}~\bibnamefont {Hemmi}}, \bibinfo
  {author} {\bibfnamefont {Z.~A.}\ \bibnamefont {Dreger}}, \bibinfo {author}
  {\bibfnamefont {K.~A.}\ \bibnamefont {Zimmerman}}, \bibinfo {author}
  {\bibfnamefont {Y.~M.}\ \bibnamefont {Gupta}}, \bibinfo {author}
  {\bibfnamefont {D.~H.}\ \bibnamefont {Torchinsky}}, \ and\ \bibinfo {author}
  {\bibfnamefont {K.~A.}\ \bibnamefont {Nelson}},\ }\bibfield  {title}
  {\enquote {\bibinfo {title} {Second-order elastic constants of
  pentaerythritol tetranitrate and cyclotrimethylene trinitramine using
  impulsive stimulated thermal scattering},}\ }\href@noop {} {\bibfield
  {journal} {\bibinfo  {journal} {J. Appl. Phys.}\ }\textbf {\bibinfo {volume}
  {104}},\ \bibinfo {pages} {073517} (\bibinfo {year} {2008})}\BibitemShut
  {NoStop}%
\bibitem [{\citenamefont {Bolme}\ and\ \citenamefont {Ramos}(2014)}]{11}%
  \BibitemOpen
  \bibfield  {author} {\bibinfo {author} {\bibfnamefont {C.~A.}\ \bibnamefont
  {Bolme}}\ and\ \bibinfo {author} {\bibfnamefont {K.~J.}\ \bibnamefont
  {Ramos}},\ }\bibfield  {title} {\enquote {\bibinfo {title} {The elastic
  tensor of single crystal rdx determined by brillouin spectroscopy},}\
  }\href@noop {} {\bibfield  {journal} {\bibinfo  {journal} {J. Appl. Phys.}\
  }\textbf {\bibinfo {volume} {116}},\ \bibinfo {pages} {183503} (\bibinfo
  {year} {2014})}\BibitemShut {NoStop}%
\bibitem [{\citenamefont {Connick}\ and\ \citenamefont
  {May}(1969{\natexlab{a}})}]{12}%
  \BibitemOpen
  \bibfield  {author} {\bibinfo {author} {\bibfnamefont {W.}~\bibnamefont
  {Connick}}\ and\ \bibinfo {author} {\bibfnamefont {F.}~\bibnamefont {May}},\
  }\bibfield  {title} {\enquote {\bibinfo {title} {Dislocation etching of
  cyclotrimethylene trinitramine crystals},}\ }\href@noop {} {\bibfield
  {journal} {\bibinfo  {journal} {J. Cryst. Growth}\ }\textbf {\bibinfo
  {volume} {5}},\ \bibinfo {pages} {65--69} (\bibinfo {year}
  {1969}{\natexlab{a}})}\BibitemShut {NoStop}%
\bibitem [{\citenamefont {Halfpenny}, \citenamefont {Roberts},\ and\
  \citenamefont {Sherwood}(1984)}]{13}%
  \BibitemOpen
  \bibfield  {author} {\bibinfo {author} {\bibfnamefont {P.}~\bibnamefont
  {Halfpenny}}, \bibinfo {author} {\bibfnamefont {K.}~\bibnamefont {Roberts}},
  \ and\ \bibinfo {author} {\bibfnamefont {J.}~\bibnamefont {Sherwood}},\
  }\bibfield  {title} {\enquote {\bibinfo {title} {{Dislocations in energetic
  materials: IV. The crystal growth and perfection of cyclotrimethylene
  trinitramine (RDX)}},}\ }\href@noop {} {\bibfield  {journal} {\bibinfo
  {journal} {J. Cryst. Growth}\ }\textbf {\bibinfo {volume} {69}},\ \bibinfo
  {pages} {73--81} (\bibinfo {year} {1984})}\BibitemShut {NoStop}%
\bibitem [{\citenamefont {Sakano}\ \emph {et~al.}(2018)\citenamefont {Sakano},
  \citenamefont {Hamilton}, \citenamefont {Islam},\ and\ \citenamefont
  {Strachan}}]{Sakano2018}%
  \BibitemOpen
  \bibfield  {author} {\bibinfo {author} {\bibfnamefont {M.}~\bibnamefont
  {Sakano}}, \bibinfo {author} {\bibfnamefont {B.}~\bibnamefont {Hamilton}},
  \bibinfo {author} {\bibfnamefont {M.~M.}\ \bibnamefont {Islam}}, \ and\
  \bibinfo {author} {\bibfnamefont {A.}~\bibnamefont {Strachan}},\ }\bibfield
  {title} {\enquote {\bibinfo {title} {{Role of molecular disorder on the
  reactivity of RDX}},}\ }\href@noop {} {\bibfield  {journal} {\bibinfo
  {journal} {J. Phys. Chem. C}\ }\textbf {\bibinfo {volume} {122}},\ \bibinfo
  {pages} {27032--27043} (\bibinfo {year} {2018})}\BibitemShut {NoStop}%
\bibitem [{\citenamefont {Connick}\ and\ \citenamefont
  {May}(1969{\natexlab{b}})}]{Connick1969}%
  \BibitemOpen
  \bibfield  {author} {\bibinfo {author} {\bibfnamefont {W.}~\bibnamefont
  {Connick}}\ and\ \bibinfo {author} {\bibfnamefont {F.~G.~J.}\ \bibnamefont
  {May}},\ }\bibfield  {title} {\enquote {\bibinfo {title} {{Dislocation
  etching of cyclotrimethylene trinitramine crystals}},}\ }\href@noop {}
  {\bibfield  {journal} {\bibinfo  {journal} {J. Cryst. Growth}\ }\textbf
  {\bibinfo {volume} {5}},\ \bibinfo {pages} {65--69} (\bibinfo {year}
  {1969}{\natexlab{b}})}\BibitemShut {NoStop}%
\bibitem [{\citenamefont {Kelly}\ and\ \citenamefont
  {Groves}(1970)}]{crystalbook}%
  \BibitemOpen
  \bibfield  {author} {\bibinfo {author} {\bibfnamefont {A.}~\bibnamefont
  {Kelly}}\ and\ \bibinfo {author} {\bibfnamefont {G.~W.}\ \bibnamefont
  {Groves}},\ }in\ \href@noop {} {\emph {\bibinfo {booktitle} {Crystallography
  and Crystal Defects}}}\ (\bibinfo  {publisher} {Longman},\ \bibinfo {year}
  {1970})\ Chap.~\bibinfo {chapter} {4}\BibitemShut {NoStop}%
\end{thebibliography}%

\end{document}